\documentclass[a4paper,11pt]{article}
\usepackage[latin1]{inputenc}
\usepackage{indentfirst}

\usepackage{amsfonts}
\usepackage{float}
\usepackage{latexsym,amssymb,amsmath}
\usepackage[dvips]{graphics,graphicx,epsfig,color}
\usepackage[lofdepth,lotdepth]{subfig}
\begin{document}
\thispagestyle{empty}
\title{\bf Computational parameter retrieval approach to the dynamic homogenization of a periodic array of rigid rectangular blocks}
\author{Armand Wirgin\thanks{LMA, CNRS, UMR 7031, Aix-Marseille Univ, Centrale Marseille, F-13453 Marseille Cedex 13, France, ({\tt wirgin@lma.cnrs-mrs.fr})} }
\date{\today}
\maketitle
\begin{abstract}
We propose to homogenize a periodic (along one direction) structure, first in order to verify the quasi-static prediction of its response to an acoustic wave arising from mixing theory, then to address the question of what becomes of this prediction at higher frequencies. This homogenization is treated as an inverse (parameter retrieval) problem, i.e., by which we: (1) generate far-field (i.e., specular reflection and transmission coefficients) response data for the given periodic structure, (2) replace (initially by thought) this (inhomgoeneous) structure by a homogeneous (surrogate) layer, (3) compute the response of the surrogate layer response for various trial constitutive properties, (4) search for the global minimum of the discrepancy between  the response data of the given structure and the various trial parameter responses (5) attribute the homogenized properties of the  surrogate layer for which the minimum of the discrepancy is attained.  We  carry out this  five-step procedure for a host of given structures  and solicitation parameters,  notably the frequency. The result is that: (i) at low frequencies and/or large filling factors, the effective constitutive properties are close to their static equivalents, i.e., the effective mass density is the product of a factor related to the given structure filling factor with the mass density of a generic substructure of the given structure and the effective velocity is equal to the  velocity in the said generic substructure, 2) at higher frequencies and/or smaller city filling factors, the effective constitutive properties are dispersive and do not take on a simple mathematical form, with this dispersion compensating for the discordance between the ways the inhomogeneous  given structure and the homogeneous surrogate layer respond to the acoustic wave. At  low frequencies, the given structure, represented as a layer  with static homogenized properties, is quite adequate to account for the principal features of the far-field response (even for the phenomenon of total transmission when the spaces between blocks are very narrow). The model of the layer with dispersive homogenized properties is more suitable to account for such features as resonances due to the excitation of surface wave modes.
\end{abstract}
\newline
\newline
Abbreviated title: Dynamic homogenization of a grating
\newline
\newline
Corresponding author: Armand Wirgin, \\ e-mail: wirgin@lma.cnrs-mrs.fr
\newpage
\tableofcontents
\newpage
\newpage
\section{Introduction}\label{intro}
The scheme by which  a solicited (by some sort of wave) inhomogeneous (porous, inclusions within a homogeneous host, etc.) medium is reduced (with regard to its response to the solicitation) to a surrogate homogeneous medium is frequently termed  'homogenization'. What is meant by homogenization also includes the manner in which the physical characteristics of the surrogate are related to the structural and physical characteristics of the original, this being  often accomplished by theoretical multiple scattering field averaging techniques \cite{ua49,wt61,sa80,lm06,sp06,tsd06,aa07,cn09} or multiscale techniques (which actually also involve field averaging) \cite{fb05,pa06,mcc17}.  Usually, the latter two techniques cannot be fully-implemented for other than static, or at least low frequency, solicitations \cite{si02,fb05,gpa16}. There exist  alternatives to theoretical  field-averaging and multiscaling, applicable to a range of (usually-low) frequencies,  which can be called:  'computational field averaging' \cite{cm12} and 'computational  parameter retrieval' \cite{wg92,qk11} approaches to dynamic homogenization.

Of course, the dynamic nature of homogenization is a central issue in many applications, such as those dealing with sound absroption,  and is therefore worthy of attention in a theoretical study.

Recent research on metamaterials \cite{la13,lc04,pm04,fb05,fa07,mt10,du12,fp15,rm15,bmm15,bpa15,bua17}  has spurred renewed interest in homogenization techniques \cite{la13,ss02,sv05,fx06,th06,tsd06,aml10,st10,csc13,gr15,ga16,mcc17}, the underlying issue being how to design an inhomogeneous medium so that it responds in a given manner (enhanced absorption \cite{gr15,gpa16,jrg17},  total transmission \cite{ll07}, reduced broadband transmission \cite{lg13,lg16} or various patterns of scattering \cite{wi78, wi88,jc17}) to a wavelike (acoustic \cite{ll07,jrg17}, elastic \cite{pa06}, optical \cite{wi78,wi88}, microwave \cite{km89}, waterwave \cite{ml99})  solicitation, the frequencies of which can exceed the quasi-static regime.  This design problem is, in fact, an inverse problem that is often solved by encasing a specimen of the medium in  a flat layer, treating the latter as if it were homogeneous, and obtaining its constitutive properties (in explicit manner in the NRW technique \cite{nr70,we74,bdp10,ss02,cg04,sv05,go10,la12,la14,wi16b} from the complex amplitudes of the reflected and transmitted waves that constitute the response  of the layer (i.e., the data) to a (normally-incident in the NRW technique) plane body wave. To the question of why encase the specimen within a flat layer  it can be answered that a simple, closed form solution is available for the associated forward problem of the reflected and transmitted-wave response of homogeneous layer to an incident plane body wave (the same being true as regards a stack of horizontal plane-parallel homogeneous layers employed in the geophysical context). Moreover, this solution is valid for all frequencies, layer thicknesses, and even incident angles and polarizations, such as is required in the homogenization problem since one strives to obtain a homogenized medium whose thickness does not depend on the specimen thickness (otherwise, how qualify it as being a medium?), but, on the contrary, he wants to find out how its properties depend on the characteristics of the solicitation. Moreover, the requirement of normal incidence (proper to the NRW technique) can be relaxed provided one accepts to search for the solution of the inverse problem in an optimization, rather than explicit, manner \cite{qk11,wi16b}.

There exist two important differences between the application-oriented metamaterial design problem and the present theoretically-oriented homogenization problem. The first difference is that our data is not design data but rather what would be obtained if the given structure were submitted to an acoustic   solicitation and the data  (a quantity related to the acoustic field) were either observed or accurately simulated. The second difference is that the NRW technique, which is widely-used in the metamaterial context, cannot be applied as such to our homogenization problem because of our requirement of angle-of-incidence diversity, which means that we must solve our inverse problem in an optimization (i.e., retrieval) manner \cite{wi16b}.

Thus, to resume what will be done hereafter:, we shall (1) generate far-field response data for a given periodic structure, (2) replace this structure by a homogeneous (surrogate) layer, (3) compute the response of the surrogate layer response for various trial constitutive properties, (4) search for the global minimum of the discrepancy between  the response data of the given structure and the various trial parameter responses of the surrogate layer (5) attribute the homogenized properties of the given structure to the surrogate layer for which the minimum of the discrepancy is attained. Furthermore,  we carry out the aforementioned five-step procedure for a host of given structures  and solicitation parameters,  notably the frequency \cite{qk11,wi16b}, which is why we refer to the whole task as 'dynamic homogenization'. This is practically feasible only because the forward problem, associated with the layer-like surrogate, possesses a simple, explicit solution at all frequencies (however, the corresponding inverse problem is strongly nonlinear and does not possess such a simple, explicit solution).
\section{Components of the inverse problem}\label{comp}
%
\subsection{Generalities}\label{gen}
A material body $\mathcal{B}(\mathbf{b})$ (herein the periodic array of blocks), subjected to a wavelike solicitation $\mathcal{S}(\mathbf{s})$, (herein a homogeneous acoustic plane wave) gives rise to the observable response $\mathcal{R}(\mathbf{b},\mathbf{s},\mathbf{r})$, with $\mathbf{b}$ and $\mathbf{s}$ denumerable sets of the physical and geometrical properties of $\mathcal{B}$ and $\mathcal{S}$ respectively and $\mathbf{r}$ the properties of the receiving (i.e., of the response field) scheme. Another material body $\mathfrak{B}$ (a somewhat different (from $\mathcal{B}$) structure), subjected to the  wavelike solicitation $\mathfrak{S}(\mathbf{S})$ (again a homogeneous plane wave),  gives rise to the response $\mathfrak{R}(\mathbf{B},\mathbf{S},\mathbf{R})$, with $\mathbf{B}$ and $\mathbf{S}$ denumerable sets of the physical and geometrical properties of $\mathfrak{B}$ and $\mathfrak{S}$ respectively and $\mathbf{R}$ the properties of the receiving scheme. The inverse problem  addressed herein is to find  the physical properties of $\mathfrak{B}$ (termed the surrogate of $\mathcal{B}$) such that the discrepancy between the responses $\mathfrak{R}$ and $\mathcal{R}$ be minimal (in the NRW technique, this difference is taken to be zero) in some sense.

The  solution of this inverse problem is denoted by $\mathbf{B}=\mathbf{\widetilde{B}}$, assuming that the solicitation $\mathfrak{S}$ is identical to $\mathcal{S}$, the receiving schemes involved in the two responses are the same, and the surrounding medium ('site' for short) of $\mathfrak{B}$ is identical to that of $\mathcal{B}$.

In this study (contrary to those of \cite{fo05,go10}) the response $\mathcal{R}$ is not actually observed (in the laboratory or in situ) but rather simulated (i.e., computed, but still denoted by $\mathcal{R}$) by means of the (so-called data) model  $\mathcal{M}$  which is the mathematical-numerical apparatus required to obtain an accurate solution $\mathcal{R}$ of a {\it forward-scattering problem}.

A second (so-called retrieval) model $\mathfrak{M}$ is the mathematical-numerical apparatus required to obtain the solution $\mathfrak{R}$ of the other forward-scattering problem connected with $\mathfrak{B}$. Since the  structure of $\mathfrak{B}$ is chosen to be simpler than that of $\mathcal{B}$, $\mathfrak{M}$ is simpler than $\mathcal{M}$. In fact, we choose the structure of the surrogate body $\mathfrak{B}$ in such a way that the forward-scattering problem connected with $\mathfrak{M}$ be solvable in simple, explicit manner. However, the solution obtained via $\mathcal{M}$ cannot be obtained in simple, explicit manner.
\subsection{Specifics}\label{specif}
The site of $\mathcal{B}$ (and $\mathfrak{B}$) consists of a lower half space filled with a  homogeneous, isotropic non-lossy  fluid and an upper half space filled with another homogeneous, isotropic fluid.

The  given structure of $\mathcal{B}$ is composed of a periodic (along $x$) set of identical, perfectly-rigid  rectangular  blocks whose heights are $h$. The spaces in between successive blocks is filled with a third homogeneous, isotropic non-lossy fluid. The surrogate structure of $\mathfrak{B}$ is simply a  layer of thickness $h$ filled with a homogeneous, isotropic, non lossy material. The site in which resides the surrogate layer is the same as that of the given structure.

All the geometrical features of $\mathcal{B}$  and $\mathfrak{B}$ are assumed to not depend on $y$ and the various interfaces (between the given structure and the two half spaces as well as between the layer and the two half spaces)  are surfaces of constant $z$ and extend indefinitely along $x$ and $y$.

 As concerns $\mathcal{B}$, $\mathbf{b}=\{\boldsymbol{\rho},\bf{c},\mathbf{h}\}$, with $\boldsymbol{\rho}$ a set of real densities, $\bf{c}$ a set of real phase velocities, and $\mathbf{h}$ a set of geometrical parameters (block heights and widths, period of the superstructure). As concerns the surrogate,  $\mathbf{B}=\{\bf{R},\bf{C},\mathbf{H}\}$, with $\bf{R}$ a set of real densities, $\bf{C}$ a set of real phase velocities, and $\mathbf{H}$ a set of geometrical parameters (layer thickness).

 Since the solicitation is, for the two configurations, a homogeneous compressional (i.e., acoustic) plane wave whose  wavevector $\mathbf{k}^{i}$ lies in the $x-z$ (sagittal) plane)), it follows that $\mathbf{s}=\mathbf{S}=\{\theta^{i},A^{i},\omega\}$, wherein $\theta^{i}$ is the incident angle, $A^{i}$ an amplitude, $\omega=2\pi f$ the angular frequency ($f$ the frequency which will be varied). The parameter sets $\boldsymbol{\rho},\bf{c}$ are assumed to not depend on $f$ (nor, of course on $\theta^{i}$  and $A^{i}$).   It is hoped that the parameter sets  $\bf{\tilde{R}},\bf{\tilde{C}}$  depend neither on $\theta^{i}$ nor on $h$ (in addition, they should not depend on $A^{i}$). It can be expected that $\bf{\tilde{R}},\bf{\tilde{C}}$ will be functions of $\boldsymbol{\rho},\bf{c}$. An open question is to what extent $\bf{\tilde{R}},\bf{\tilde{C}}$ depend on $f$.

 Since the receivers are, for the two configurations, a set of idealized transducers situated very far from, and on both sides of the given structure or its surrogate, it follows that $\mathbf{r}=\mathbf{R}=\{\theta^{-}_{1},\theta^{+}_{1},\theta^{-}_{2},\theta^{+}_{2},...,\theta^{-}_{n_{r}},\theta^{+}_{n_{r}}\}$, with $\theta^{-}_{j}$ designating the angle of  reflection in the lower half space and $\theta^{t}_{+}$ the angle of emergence of the transmitted wave in the upper half space, with  $n_{r}$ the number of receiver pairs.

\begin{figure}[ptb]
\begin{center}
\includegraphics[width=10cm] {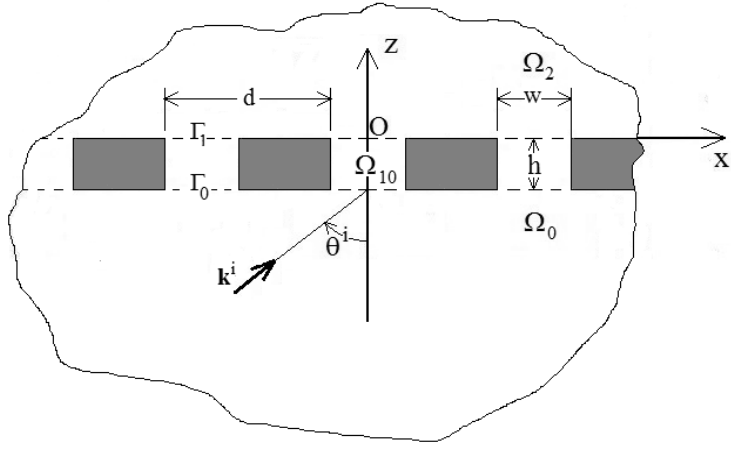}
 \caption{Sagittal plane view of the configuration assumed in data model $\mathcal{M}$. $\Omega_{0}$ is the lower half-space domain, $\Omega_{1}=\cup_{n\in\mathbb{Z}}\Omega_{1n}$  the domain occupied by the given periodic structure, $\Omega_{1n}$ the domain of the filler space of width $w$ between successive  blocks,  the blocks being of rectangular cross section  and height $h$, with $d$ the period (along $x$) of the given periodic structure, and $\Omega_{2}$ the  upper half space domain. The  interface (i.e., the line $z=-h_{2}$) between $\Omega_{0}$ and $\Omega_{1}$ is designated by $\Gamma_{0}=\cup_{n\in\mathbb{Z}}\Gamma_{0n}$ and the interface  between  $\Omega_{1}$ and $\Omega_{2}$  by $\Gamma_{1}=\cup_{n\in\mathbb{Z}}\Gamma_{1n}$, in which $\Gamma_{m0}$ is the portion of $\Gamma_{m}$ included between $x=-w/2$ and $x=x/2$.}
  \label{fig1}
  \end{center}
\end{figure}

\begin{figure}[ptb]
\begin{center}
\includegraphics[width=10cm] {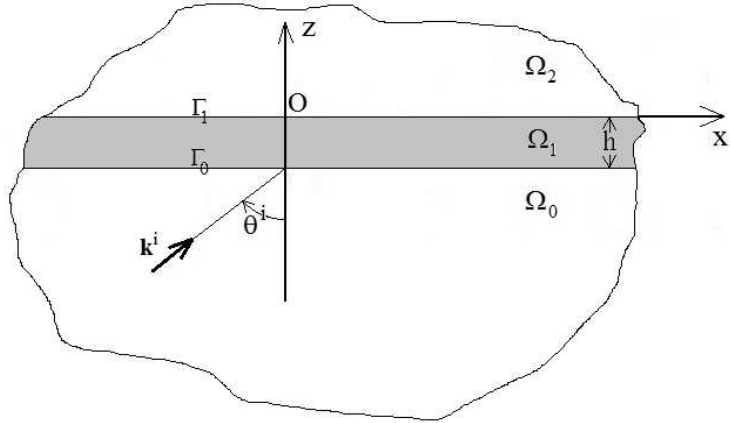}
 \caption{Sagittal plane view of the configuration assumed in retrieval model $\mathfrak{M}$. $\Omega_{0}$ is the lower  half-space domain,  $\Omega_{1}$  the  surrogate  domain in the form of a layer  of thickness $h$. $\Omega_{2}$ is the half-space above the superstructure. The  interface (i.e., the line $z=-h$) between $\Omega_{0}$ and $\Omega_{1}$ is designated by $\Gamma_{0}$  and the interface  between  $\Omega_{1}$ and $\Omega_{2}$  by $\Gamma_{1}$.}
  \label{fig2}
  \end{center}
\end{figure}

 The assumed  $y$-independence of the acoustic solicitation, as well as the fact that it was assumed that the blocks of the given structure and the layer of the surrogate structure do not depend on $y$ entails that all the acoustic field functions do not depend on $y$. Thus, the two forward problems associated with $\mathcal{M}$ and $\mathfrak{M}$ are both 2D, so that the analysis takes place in the sagittal plane in which the vector joining the origin O to a point whose coordinates are $x,z$ is denoted by $\mathbf{x}$ (see figs. \ref{fig1}-\ref{fig2}).

Let $f(\mathbf{b},\mathbf{s},\mathbf{r}|\omega)$ designate a wavefield response function at angular frequency $\omega$ to the forward-scattering problem of model $\mathcal{M}$ and $F(\mathbf{B},\mathbf{S}=\mathbf{s},\mathbf{R}=\mathbf{r}|\omega)$ the corresponding wavefield response function at angular frequency $\omega$ to the forward-scattering problem of model $\mathfrak{M}$. The chosen discrepancy functional (called cost functional $\kappa$ hereafter) is simply related to the difference of these  pairs of  wavefield response functions at $n_{r}$ pairs of receiver angles  $\{\theta^{-}_{1},\theta^{+}_{1},\theta^{-}_{2},\theta^{+}_{2},...,r_{n_{r}}^{-}=\theta^{-}_{n_{r}},r_{n_{r}}^{+}=\theta^{+}_{n_{r}}\}$,  and is given by
\begin{equation}\label{0-010}
\kappa(\mathbf{b},\mathbf{B},\mathbf{s},\mathbf{r}|\omega)=\frac{\sum_{j=1}^{n_{r}}\left(\|F(\mathbf{B},\mathbf{s},r_{j}^{-}|\omega)-
f(\mathbf{b},\mathbf{s},r_{j}^{-}|\omega)\|^{2}+\|F(\mathbf{B},\mathbf{s},r_{j}^{+}|\omega)-
f(\mathbf{b},\mathbf{s},r_{j}^{+}|\omega)\|^{2}\right)}
{\sum_{j=1}^{n_{r}}\left(\|f(\mathbf{b},\mathbf{s},r_{j}^{-}|\omega)\|^{2}+\|f(\mathbf{b},\mathbf{s},r_{j}^{+}|\omega)\|^{2}\right)}~.
\end{equation}
Recall that $\mathbf{b}$, $\mathbf{s}$, $\mathbf{r}$ are fixed, which is not the case for $\mathbf{B}$ because the idea is to vary it so as to minimize $\kappa$. This variation of $\mathbf{B}$ is carried out until the attainment of the global minimum of $\kappa$ at which moment  $\mathbf{B}$ is said to  equal $\mathbf{\tilde{B}}$ corresponding to the solution of the inverse problem at frequency $\omega$. Subsequently, $\omega$ can be varied so as to find out to what extent $\mathbf{\tilde{B}}$ depends on $\omega$.
%
%
%
\newpage
Of principal interest in the synthesis or homogenization context is how $\mathbf{\tilde{B}}$ is functionally-related to $\mathbf{b}$. This can be determined empirically or by regression analysis after varying $\mathbf{b}$ and plotting the various entries of $\mathbf{\tilde{B}}$ versus those of $\mathbf{b}$.
\section{Data obtained by the resolution of the forward-scattering problem  corresponding to model $\mathcal{M}$}
The data is related to the pressure field in the far-field region of  both the lower and upper half-spaces of the configuration involving $\mathcal{B}$ in fig. \ref{fig1}. Implicit in the problem formulation is a domain decomposition (DD) ($\mathbb{R}^{2}$ in terms of $\Omega_{l}~;~l=0,1,2$), each subdomain being associated with a particular pressure field representation.
\subsection{The boundary-value problem}\label{bvp1}
The real phase velocities in $\Omega_{l}; l=0,2$ are $c^{[l]}$, and the real phase velocities in $\Omega_{1n}~n\in \mathbb{Z}$ are $c^{[1]}$. Both $c^{[j]}~l=0,1,2$  are $\ge 0$ and assumed do not depend on $\omega$.

The positive real density in $\Omega_{l}~;~l=0,2$ is $\rho^{[l]}>0$ and the real positive real densities in $\Omega_{1n}~n\in \mathbb{Z}$ are $\rho^{[1]}$.  The three densities are assumed to not depend on $\omega$.

The set $\mathbf{b}=\{\boldsymbol{\rho},\mathbf{c},\mathbf{h}\}$ is $\{\rho^{[0]},\rho^{[1]},\rho^{[2]},c^{[0]},c^{[1]},c^{[2]},w,d,h\}$ and is implicit in the field quantities (except the incident wavefield) referred-to hereafter. The sets $\mathbf{s}$ and $\mathbf{r}$ will also be considered to be implicit in the wavefields that appear in the following.

The  wavevector $\mathbf{k}^{i}$  is of the form $\mathbf{k}^{i}=(k_{x}^{i},k_{z}^{i})=(k^{[0]}\sin\theta^{i},k^{[0]}\cos\theta^{i})$ wherein  $\theta^{i}$ is the angle of incidence (see fig. \ref{fig1}), and $k^{[j]}=\omega/c^{[j]}$.

The total pressure wavefield $u(\mathbf{x},\omega)$ in $\Omega_{l}~;~l=0,1,2$ is designated by $u^{[l]}(\mathbf{x},\omega)$.  The incident wavefield is
\begin{equation}\label{1-000}
u^{[0]+}(\mathbf{x},\omega)=u^{i}(\mathbf{x},\omega)=A^{[0]+}(\omega)\exp[i(k_{x}^{i}x+k_{z}^{i}z)]~,
\end{equation}
wherein $A^{[0]+}(\omega)=A^{i}\sigma(\omega)$ and $\sigma(\omega)$ is the spectrum  of the solicitation.

The plane wave nature of the solicitation and the $d$-periodicity of $\Gamma_{0}$ and $\Gamma_{1}$  entails the quasi-periodicity of the field, whose expression is the Floquet condition
\begin{equation}\label{1-005}
u(x+d,z,\omega)=u(x,z,\omega)\exp(ik_{x}^{i}d)~;~\forall\mathbf{x}\in \Omega_{0}+\Omega_{1}+\Omega_{2}~.
\end{equation}
Consequently, as concerns the response in $\Omega_{1}$, it suffices to examine the field in $\Omega_{10}$.

The boundary-value problem in the space-frequency domain translates to the following relations (in which the superscripts $+$ and $-$ refer to the upgoing and downgoing  waves respectively) satisfied by the total displacement field $u^{[l]}(\mathbf{x};\omega)$ in $\Omega_{l}$:
\begin{equation}\label{1-010}
u^{[l]}(\mathbf{x},\omega)=u^{[l]+}(\mathbf{x},\omega)+u^{[l]-}(\mathbf{x},\omega)~;~l=0,1,2~,
\end{equation}
\begin{equation}\label{1-020}
u_{,xx}^{[l]}(\mathbf{x},\omega)+u_{,zz}^{[l]}(\mathbf{x},\omega)+(k^{[l]})^{2}u^{[l]}(\mathbf{x},\omega)=0~;~\mathbf{x}\in \Omega_{l}~;~l=0,1,2~.
\end{equation}
\begin{equation}\label{1-030}
(\rho^{[0]})^{-1}u_{,z}^{[0]}(x,-h,\omega)=0~;~\forall x\in [-d/2,w/2]\cup [w/2,d/2]~,
\end{equation}
\begin{equation}\label{1-033}
(\rho^{[2]})^{-1}u_{,z}^{[2]}(x,0,\omega)=0~;~\forall x\in [-d/2,w/2]\cup [w/2,d/2]~,
\end{equation}
\begin{equation}\label{1-035}
(\rho^{[2]})^{-1}u_{,x}^{[1]}(\pm w/2,z,\omega)=0~;~\forall z\in [0,h_{2}]~,
\end{equation}
\begin{equation}\label{1-040}
u^{[l]}(\mathbf{x},\omega)-u^{[1+1]}(\mathbf{x},\omega)=0~;~\mathbf{x}\in \Gamma_{l}~;~l=0,1~,
\end{equation}
\begin{equation}\label{1-050}
(\rho^{[l]})^{-1}u_{,z}^{[l]}(\mathbf{x},\omega)-(\rho^{[l+1]})^{-1}u_{,z}^{[1+1]}(\mathbf{x},\omega)=0~;~\mathbf{x}\in \Gamma_{l}~;~l=0,1~,
\end{equation}
wherein   $u_{,\zeta}$ ($u_{,\zeta\zeta}$) denotes the first (second) partial derivative of $u$ with respect to $\zeta$. Eq. (\ref{1-020}) is the space-frequency  wave equation for compressional sound waves, (\ref{1-030})-(\ref{1-035}) the rigid-surface  boundary conditions, (\ref{1-040}) the expression of continuity of pressure across the two interfaces $\Gamma_{0}$ and $\Gamma_{0}$  and (\ref{1-050}) the expression of continuity of normal velocity across these same interfaces.

Since $\Omega_{l}~;~l=0,2$ are of half-infinite extent, the field therein must obey the radiation conditions
\begin{equation}\label{1-060}
u^{[l]-}(\mathbf{x},\omega)\sim \text{outgoing waves}~;~\mathbf{x}\rightarrow\infty~;~l=0,2~.
\end{equation}
Various (usually integral equation) rigorous approaches \cite{mi49,km89,ml99} have been employed to solve this boundary value problem. Herein, we outline another rigorous technique, based on the domain decomposition and  separation of variables technique previously developed in \cite{wi78,wi88}.
\subsection{Field representations via separation of variables (DD-SOV)}\label{sov1}
The application of  the domain decomposition-separation of variables (DD-SOV) technique, The Floquet condition, and the radiation conditions gives rise, in the two half-spaces, to the field representations:
\begin{equation}\label{2-010}
u^{[0]}(\mathbf{x},\omega)=\sum_{n\in\mathbb{Z}}\left(A_{0}^{[l]+}(\omega)\exp[i(k_{xn}x+ k_{zn}^{[0]}z)]+
A_{0}^{[l]-}(\omega)\exp[i(k_{xn}x- k_{zn}^{[0]}z)]\right)~,
\end{equation}
\begin{equation}\label{2-011}
u^{[2]}(\mathbf{x},\omega)=\sum_{n\in\mathbb{Z}}A_{n}^{[2]+}(\omega)\exp[i(k_{xn}x+ k_{zn}^{[2]}z)]~,
\end{equation}
wherein:
\begin{equation}\label{2-012}
k_{xn}=k_{x}^{i}+\frac{2n\pi}{d}~,
\end{equation}
\begin{equation}\label{2-020}
k_{zn}^{[l]}=\sqrt{\big(k^{[l]}\big)^{2}-(k_{xn})^{2}}~~;~~\Re k_{zn}^{[l]}\ge 0~~,~~\Im k_{zn}^{[l]}\ge 0~~\omega>0~,
\end{equation}
and, on account of (\ref{1-000}),
\begin{equation}\label{2-030}
A_{n}^{[0]+}(\omega)=A^{[0]+}(\omega)~\delta_{n0}~,
\end{equation}
with $\delta_{n0}$ the Kronecker delta symbol.
In the central block, the DD-SOV technique, together with the rigid body boundary condition (\ref{1-035}, lead to
\begin{equation}\label{2-040}
u^{[1]}(\mathbf{x},\omega)=\sum_{m=0}^{\infty}\left(A_{m}^{[1]+}(\omega))\exp[iK_{zm}^{[1]}z)]+
A_{m}^{[1]-}(\omega))\exp[-iK_{zm}^{[1]}z)]\right)\cos[K_{xm}(x+w/2)]~,
\end{equation}
in which
\begin{equation}\label{2-050}
K_{xm}=\frac{m\pi}{w}~,
\end{equation}
\begin{equation}\label{2-060}
K_{zm}^{[1]}=\sqrt{\big(k^{[1]}\big)^{2}-(K_{xm})^{2}}~~;~~\Re K_{zm}^{[1]}\ge 0~~,~~\Im K_{zm}^{[1]}\ge 0~~\omega>0~.
\end{equation}
%
\subsection{Expressions for each of the four sets of unknowns}
From the remaining boundary and continuity conditions it ensues the four sets of relations:
\begin{equation}\label{4-000}
(\rho^{[0]})^{-1}\int_{-d/2}^{d/2}u_{,z}^{[0]}(x,-h,\omega)\exp(-ik_{xj}\frac{dx}{d}=
(\rho^{[1]})^{-1}\int_{-d/2}^{d/2}u_{,z}^{[1]}(x,-h,\omega)\exp(-ik_{xj}\frac{dx}{d}~;~\forall j\in \mathbb{Z}~,
\end{equation}
\begin{equation}\label{4-005}
(\rho^{[2]})^{-1}\int_{-d/2}^{d/2}u_{,z}^{[2]}(x,0,\omega)\exp(-ik_{xj}\frac{dx}{d}=
(\rho^{[1]})^{-1}\int_{-d/2}^{d/2}u_{,z}^{[1]}(x,0,\omega)\exp(-ik_{xj}\frac{dx}{d}~;~\forall j\in \mathbb{Z}~,
\end{equation}
\begin{multline}\label{4-010}
\int_{-w/2}^{w/2}u^{[0]}(x,-h,\omega)\cos[K_{xl}(x+w/2)]\frac{dx}{w/2}=\\
\int_{-w/2}^{w/2}u^{[1]}(x,-h,\omega)\cos[K_{xl}(x+w/2)]\frac{dx}{w/2}
~;~l=0,1,2,...~,
\end{multline}
\begin{multline}\label{4-015}
\int_{-w/2}^{w/2}u^{[2]}(x,0,\omega)\cos[K_{xl}(x+w/2)]\frac{dx}{w/2}=\\
\int_{-w/2}^{w/2}u^{[1]}(x,0,\omega)\cos[K_{xl}(x+w/2)]\frac{dx}{w/2}
~;~l=0,1,2,...~,
\end{multline}
which should suffice to determine the four sets of unknown coefficients $\{A^{[0]-}\}$, $\{A^{[1]+}\}$, $\{A^{[1]-}\}$, and $\{A^{[2]+}\}$. Employing the DD-SOV field representations in these four relations gives rise to:
\begin{equation}\label{4-020}
A_{j}^{[0]-}=A_{j}^{[0]+}\big(e_{j}^{-}\big)^{-2}-
\frac{\rho^{[0]}w}{2\rho^{[1]}d}\frac{e_{j}^{-}}{k_{zj}^{[0]}}
\sum_{m=0}^{\infty}\left[A_{m}^{[1]+}e_{m}^{-}-A_{m}^{[1]-}e_{m}^{+}\right]K_{zm}^{[1]}E_{jm}^{-}
~;~\forall j\in\mathbb{Z}~,
\end{equation}
\begin{equation}\label{4-030}
A_{j}^{[2]+}=\frac{\rho^{[2]}w}{2\rho^{[1]}d}\frac{1}{k_{zj}^{[2]}}
\sum_{m=0}^{\infty}\left[A_{m}^{[1]+}-A_{m}^{[1]-}\right]K_{zm}^{[1]}E_{jm}^{-}~;~\forall j\in\mathbb{Z}~,
\end{equation}
\begin{equation}\label{4-040}
A_{l}^{[1]+}e_{l}^{-}+A_{l}^{[1]-}e_{l}^{+}=\frac{\epsilon_{l}}{2}
\sum_{n\in\mathbb{Z}}\left[A_{n}^{[0]+}\varepsilon_{n}^{-}+A_{n}^{[0]-}\varepsilon_{n}^{+}\right]E_{nl}^{+}~;~l=0,1,2,..~,
\end{equation}
\begin{equation}\label{4-050}
A_{l}^{[1]+}+A_{l}^{[1]-}=\frac{\epsilon_{l}}{2}
\sum_{n\in\mathbb{Z}}A_{n}^{[2]+}E_{nl}^{+}~;~l=0,1,2,..~,
\end{equation}
wherein:
\begin{equation}\label{4-060}
e_{m}^{\pm}=\exp(\pm iK_{zm}^{[1]}h)~,~\varepsilon_{n}^{\pm}=\exp(\pm ik_{zn}h)~,~E_{nm}^{\pm}=\int_{-w/2}^{w/2}\exp(\pm ik_{zn}x)\cos[K_{xm}(x+w/2)]\frac{dx}{w/2}
\end{equation}
and $\epsilon_{0}=1$, $\epsilon_{m>0}=2$. More specifically:
\begin{equation}\label{4-070}
E_{nm}^{\pm}=i^{m}\text{sinc}([\pm k_{xn}+K_{xm}]w/2)+i^{-m}\text{sinc}([\pm k_{xn}-K_{xm}]w/2)
\end{equation}
in which $\text{sinc}(\zeta)=\sin(\zeta)/\zeta$ and $\text{sinc}(0)=1$.

We can go a step further by inserting two of the expressions (\ref{4-020})-(\ref{4-050}) into the other two and interchange summations to obtain:
\begin{equation}\label{4-080}
\begin{array}{c}
\sum_{m=0}^{\infty}\left(P_{lm}^{11}Q_{m}^{1}+P_{lm}^{12}Q_{m}^{2}\right)=R_{l}\\\\
\sum_{m=0}^{\infty}\left(P_{lm}^{21}Q_{m}^{1}+P_{lm}^{22}Q_{m}^{2}\right)=R_{2}
\end{array}
~;~l=0,2,...~,
\end{equation}
in which
\begin{equation}\label{4-090}
P_{lm}^{11}=\delta_{lm}+\frac{\epsilon_{l}w\rho^{[0]}}{4d\rho^{[1]}}e_{l}^{+}e_{m}^{-}K_{zm}^{[1]}S^{[0]}_{lm}~,~
P_{lm}^{12}=\left(e_{l}^{+}\right)^{2}\delta_{lm}-\frac{\epsilon_{l}w\rho^{[0]}}{4d\rho^{[1]}}e_{l}^{+}e_{m}^{+}K_{zm}^{[1]}S^{[0]}_{lm}
~,~R_{m}=A^{[0]+}\epsilon_{l}e_{l}^{+}\varepsilon_{0}^{-}E_{0l}^{+}
\end{equation}
\begin{equation}\label{4-100}
P_{lm}^{21}=\delta_{lm}-\frac{\epsilon_{l}w\rho^{[2]}}{4d\rho^{[1]}}e_{l}^{+}e_{m}^{-}K_{zm}^{[1]}S^{[2]}_{lm}~,~
P_{lm}^{22}=\delta_{lm}+\frac{\epsilon_{l}w\rho^{[2]}}{4d\rho^{[1]}}K_{zm}^{[1]}S^{[0]}_{lm}
~,~R_{m}=0~,~S_{lm}^{[j]}=\sum_{n\in\mathbb{Z}}\frac{E_{nl}^{+}E_{nm}^{-}}{k_{zn}^{[j]}}~,
\end{equation}
and $Q_{m}^{1}=A_{m}^{[1]+}$, $Q_{m}^{2}=A_{m}^{[1]-}$.
\subsection{Numerical issues concerning the resolution of the system of equations for $\{A_{m}^{[2]}\}$}
We strive to obtain numerically the sets $\{A_{m}^{[1]\pm}\}$ from the linear system of equations (\ref{4-080}). Once these sets are found, they are introduced into (\ref{4-020})-(\ref{4-030}) to obtain the sets $\{A_{n}^{[0]-}\}$ and $\{A_{n}^{[2]+}\}$.

Concerning the resolution of the infinite system of linear equations (\ref{4-080}), the procedure is basically to replace it by the finite system of linear equations
\begin{equation}\label{5-010}
\begin{array}{c}
\sum_{m=0}^{M}\left(P_{lm}^{11(N)}Q_{m}^{1}+P_{lm}^{12(N)}Q_{m}^{2}\right)=R_{l}\\\\
\sum_{m=0}^{M}\left(P_{lm}^{21(N)}Q_{m}^{1}+P_{lm}^{22(N)}Q_{m}^{2}\right)=R_{2}
\end{array}
~;~l=0,2,...,M~,
\end{equation}
in which $P_{lm}^{jk(N)}$ signifies that the series in $S_{lm}^{[k]}$ therein is limited to the terms $n=0,\pm 1,...,\pm N$, $N$ having been chosen to be sufficiently large to obtain numerical convergence of these series, and to increase $M$ so as to generate the sequence of numerical solutions $\{Q_{0}^{j(0)}\}$, $\{Q_{0}^{j(1)},Q_{1}^{j(1)}\}$,....until the values of the first few members of these sets stabilize and the remaining members become very small. This is usually obtained for  values of $M$, that are all the smaller  the lower is the  frequency.

When all the coefficients (we mean those whose values depart significantly from zero) are found, they enable the computation of the far-field acoustic response  that is of interest in this paper. The so-obtained numerical solutions for thee far-field coefficients, can, for all practical purposes,  be considered to be 'exact' since they compare very well with their finite element or integral equation counterparts in  \cite{mi49,km89, ml99}, and are therefore taken to be the constituents of the data required to solve the inverse problem. We give a more precise definition of this data in the next section.
\subsection{A conservation principle}\label{consv}
Using Green's second identity, it is rather straightforward to demonstrate the following conservation principle:
\begin{equation}\label{5-020}
\mathcal{F}{^-}+\mathcal{F}^{+}=1~,
\end{equation}
wherein $\mathcal{F}^{-}$ and $\mathcal{F}^{+}$ are what can be termed hemisperical reflected and transmitted fluxes respectively,  given by:
\begin{equation}\label{5-030}
\mathcal{F}^{-}=\Re\sum_{n\in\mathbb{Z}}\frac{\|A_{n}^{[0]-}\|^{2}k_{zn}^{[0]}}{\|A^{[0]+}\|^{2}k_{z0}^{[0]}}~~,~~
\mathcal{F}^{+}=\Re\sum_{n\in\mathbb{Z}}\frac{\|A_{n}^{[2]-}\|^{2}k_{zn}^{[2]}\rho^{[0]}}{\|A^{[0]+}\|^{2}k_{z0}^{[0]}\rho^{[2]}}~.
\end{equation}
These expressions show (since $k_{z0}^{[0]}=k_{z}^{i}$ is real) that the only contributions to $\mathcal{F}^{-}$ stem from diffracted-reflected waves for which $k_{zn}^{[0]}$ is real, and the only contributions to $\mathcal{F}^{+}$ stem from diffracted-transmitted waves for which $k_{zn}^{[2]}$ is real. Real $k_{zn}^{[0]}$ corresponds to homogeneous  plane waves in the lower half space and real $k_{zn}^{[2]}$ to homogeneous  plane waves in the upper half space. The angles of emergence of these observable homogeneous waves are $\theta_{n}^{-}=\arcsin(k_{xn}/k^{[0]})$ and $\theta_{n}^{+}=\arcsin(k_{xn}/k^{[2]})$ (note that $\theta_{0}^{-}=\theta^{i}$ is the angle of specular reflection in the sense of Snell, and $\theta_{0}^{+}=\arcsin(k^{[0]}\sin\theta^{i}/k^{[2]})$ the angle of refraction in the sense of Fresnel). Thus, the flux in each half space is composed of a denumerable, finite, set of  subfluxes, which fact is expressed by:
\begin{equation}\label{5-040}
\mathcal{F}^{-}=\sum_{n\in\mathbb{H}^{-}}\mathcal{F}_{n}^{-}~~,~~
\mathcal{F}^{+}=\sum_{n\in\mathbb{H}^{+}}\mathcal{F}_{n}^{+}~.
\end{equation}
wherein:
\begin{equation}\label{5-050}
 \mathcal{F}_{n}^{-}=\frac{\|A_{n}^{[0]-}\|^{2}k_{zn}^{[0]}}{\|A^{[0]+}\|^{2}k_{z0}^{[0]}}~~,~~
 \mathcal{F}_{n}^{+}=\frac{\|A_{n}^{[2]-}\|^{2}k_{zn}^{[2]}\rho^{[0]}}{\|A^{[0]+}\|^{2}k_{z0}^{[0]}\rho^{[2]}}~.
\end{equation}
and $\mathbb{H}^{-}$  the set of $n$ for which $k_{zn}^{[0]}$ is real, whereas  $\mathbb{H}^{+}$ is the set of $n$ for which $k_{zn}^{[2]}$ is real.

This shows that there are various ways in which one can define far-field (observable) data:
\begin{enumerate}
\item $A_{0}^{[0]-}$ and $A_{0}^{[2]+}$ (this is what is employed in the NRW method for $\theta^{i}=0^{\circ}$),\\
\item $A_{n}^{[0]-}~;~\forall n\in\mathbb{H}^{-}$  and $A_{n}^{[2]+}~;~\forall n\in\mathbb{H}^{-}$, (this is equivalent to the first choice at low frequencies, i.e., when only the $n=0$ plane waves are homogeneous))\\
\item $\mathcal{F}_{0}^{-}$ and $\mathcal{F}_{0}^{+}$,\\
\item  $\mathcal{F}_{n}^{-}~;~\forall n\in\mathbb{H}^{-}$  and $\mathcal{F}_{n}^{+}~;~\forall n\in\mathbb{H}^{+}$,\\
\item $\mathcal{F}^{-}$ and $\mathcal{F}^{+}$.
\end{enumerate}
In our approach to the resolution of the inverse problem, we shall employ only the first two types of response data as experience shows that the other three types of data are inadequate (i.e., do not contain enough information because of phase suppression) for this task.
\section{Response fields obtained by the resolution of the forward-scattering problem  corresponding to the surrogate model $\mathfrak{M}$}
Here we compute the total displacement field in the far-field region of  layer-like (surrogate) structure of the configuration involving $\mathfrak{B}$ in fig. \ref{fig2}.
\subsection{The boundary-value problem}
The solicitation, bottom layer and half space are as previously (i.e., in the problem corresponding to model $\mathcal{M}$). The surrogate superstructure  occupies the layer-like domain $\Omega_{1}$ (see fig. \ref{fig2}) in which the properties are designated by the superscript 1. Thus, the boundary-value problem is expressed by (\ref{1-000}), (\ref{1-010}), (\ref{1-020}), (\ref{1-040}), (\ref{1-050}), (\ref{1-060}) (in which $\rho^{[l]}$ is replaced by $R^{[l]}$),  $c^{[l]}$  by $C^{[l]}$),  and $u^{[l]}$  by $U^{[l]}$).

 In addition, since we assume that the site of $\mathfrak{B}$ is identical to that of $\mathcal{B}$:
\begin{equation}\label{7-015}
C^{[l]}=c^{[l]}~~,~~R^{[l]}=\rho^{[l]}~;~l=0,2~.
\end{equation}
%
\subsection{DD-SOV Field representations}\label{sov2}
Separation of variables and the radiation condition lead to the field representations:
\begin{equation}\label{7-017}
U^{[l]}(\mathbf{x},\omega)=U^{[l]+}(\mathbf{x},\omega)+U^{[l]-}(\mathbf{x},\omega)~;~\forall x\in \Omega_{l}~,~l=0,1,2~.
\end{equation}
with
\begin{equation}\label{7-020}
U^{[l]\pm}(\mathbf{x},\omega)=A^{[l]\pm}\exp\left[i(k_{x}x\pm k_{z}^{[l]}z)\right]~;~\forall x\in \Omega_{l}~,~l=0,1,2~.
\end{equation}
  in which $A^{[0]+}$ is as previously, and the relation to previous wavenumbers is as follows: $k^{[l]}=\omega/C^{[l]}$, $k_{x}=k_{x0}=k^{[0]}sin\theta^{i}$, $k_{z}^{[l]}=k_{z0}^{[l]}=\sqrt{\big(k^{[l]}\big)^2-\big(k_{x}\big)^2}$.
\subsection{Solutions for the plane-wave coefficients and displacement fields}
The  four interface continuity relations lead to
\begin{equation}\label{7-050}
\begin{array}{rcr}
  A^{[0]-}e^{[0]}-A^{[1]+}\big(e^{[1]}\big)^{-1}-A^{[1]-}e^{[1]} &=& -A^{[0]+}\big(e^{[0]}\big)^{-1} \\
  -A^{[0]-}e^{[0]}-A^{[1]+}g^{[10]}\big(e^{[1]}\big)^{-1}+A^{[1]-}g^{[10]}e^{[1]} &=& -A^{[0]+}\big(e^{[0]}\big)^{-1} \\
  A^{[1]+}+A^{[1]-}-A^{[2]+} &=& 0 \\
  A^{[1]+}-A^{[1]-}-A^{[2]+}g^{[21]} &=& 0
\end{array}
~,
\end{equation}
(with $e^{[j]}=\exp(ik_{z}^{[j]}h)$ and $g^{[jk]}=g^{[j]}/g^{[k]}$, $g^{[j]}=k_{z}^{[j]}/R^{[j]}$)
for the four unknowns $A^{[0]-},~A^{[1]+}$, $A^{[1]-}~A^{[2]+}$. The solution of (\ref{7-050}), for the two unknowns of interest in the far-field region, is:
\begin{equation}\label{7-060}
A^{[0]-}=
-A^{[0]+}\big(e^{[0}\big)^{-2}\left[
\big(g^{[2]}+g^{[1]}\big)\big(g^{[1]}-g^{[0]}\big)\big(e^{[1]}\big)^{-1}+
\big(g^{[2]}-g^{[1]}\big)\big(g^{[1]}+g^{[0]}\big)e^{[1]}
\right]D^{-1}~,
\end{equation}
\begin{equation}\label{7-070}
A^{[2]+}=
A^{[0]+}\big(e^{[0}\big)^{-1}
\big[4g^{[1]}g^{[0]}\big]
D^{-1}~,
\end{equation}
wherein
\begin{equation}\label{7-080}
D=\big(g^{[2]}+g^{[1]}\big)\big(g^{[1]}+g^{[0]}\big)\big(e^{[1]}\big)^{-1}+
  \big(g^{[2]}-g^{[1]}\big)\big(g^{[1]}-g^{[0]}\big)e^{[1]}~.
\end{equation}
Note that in model $\mathcal{M}$, the constitutive parameters $c^{[1]}$ and $\rho^{[1]}$ are given and therefore known, whereas in model $\mathfrak{M}$ the parameters (actually, as we shall see: functions) $C^{[1]}$ and $R^{[1]}$ are unknown. In fact, they will be varied during the inversion process so as to obtain a global minimum in the cost functional at which moment the current value of these parameters will be considered to be the solution $\tilde{C}^{[2]}$, $\tilde{R}^{[2]}$ of the inverse problem.
\subsection{A conservation principle}\label{Consv}
Again using Green's second identity, leads in rather straightforward manner to  the following conservation principle:
\begin{equation}\label{7-120}
\mathfrak{F}{^-}+\mathfrak{F}^{+}=1~,
\end{equation}
wherein $\mathfrak{F}^{-}$ and $\mathfrak{F}^{+}$ are the hemisperical= single-wave reflected and transmitted fluxes respectively,  given by:
\begin{equation}\label{7-130}
\mathfrak{F}^{-}=\frac{\|A^{[0]-}\|^{2}}{\|A^{[0]+}\|^{2}}~~,~~
\mathfrak{F}^{+}=\frac{\|A^{[2]-}\|^{2}k_{z}^{[2]}R^{[0]}}{\|A^{[0]+}\|^{2}k_{z}^{[0]}R^{[2]}}~.
\end{equation}
These expressions show, unsurprisingly (since $k_{z}^{[0]}=k_{z}^{i}$ and $k_{z}^{[2]}$ are real) that the only contributions to $\mathfrak{F}^{-}$ stems from the single specularly-reflected  homogeneous plane wave(s) and the only contributions to $\mathfrak{F}^{+}$ stems from  the single transmitted homogeneous plane wave (there exist no other transmitted) wave(s).  The angles of emergence of these observable homogeneous waves are $\theta^{-}=\arcsin(k_{x}/k^{[0]})=\theta^{i}$ and $\theta^{+}=\arcsin(k_{x}/k^{[2]})$ (note again that $\theta^{-}=\theta^{i}$ is the angle of specular reflection in the sense of Snell, and $\theta^{+}=\arcsin(k^{[0]}\sin\theta^{i}/k^{[2]})$ the angle of refraction in the sense of Fresnel).

This shows that there are two ways in which one can define far-field (observable) response:
\begin{enumerate}
\item $A^{[0]-}$ and $A^{[2]+}$ (this is what is employed in the NRW method for $\theta^{i}=0^{\circ}$),\\
\item $\mathfrak{F}^{-}$ and $\mathfrak{F}^{+}$.
\end{enumerate}
In our approach to the resolution of the inverse problem, we shall employ only the first type  of response as experience shows that the other  type of response is inadequate (i.e., do not contain enough information because of phase suppression) for this task.
\section{More on the inversion procedure}
 On account of what was brought to fore in sects. \ref{consv} and \ref{Consv}, we can write the general cost function of (\ref{0-010}) as
\begin{multline}\label{8-010}
\kappa(\mathbf{b},\mathbf{B},\mathbf{s},\mathbf{r}|\omega)=\\
\frac{
\sum_{n\in\mathbb{H}^{-}}
\|F(\mathbf{B},\mathbf{s},\theta^{-}|\omega)-f(\mathbf{b},\mathbf{s},\theta_{n}^{-}|\omega)\|^{2}
+\sum_{n\in\mathbb{H}^{+}}
\|F(\mathbf{B},\mathbf{s},\theta^{+}|\omega)-f(\mathbf{b},\mathbf{s},\theta_{n}^{+}|\omega)\|^{2}
}
{
\sum_{n\in\mathbb{H}^{-}}\|f(\mathbf{b},\mathbf{s},\theta_{n}^{-}|\omega)\|^{2}+
\sum_{n\in\mathbb{H}^{+}}\|f(\mathbf{b},\mathbf{s},\theta_{n}^{+}|\omega)\|^{2}
}~.
\end{multline}
In view of the previously-mentioned choices of response observables, we define four particular choices of $f$ and $F$ pairs:
\begin{equation}\label{9-020}
f(\mathbf{b},\mathbf{s},\theta_{n}^{\pm}|\omega)=A_{n}^{[1\pm\delta_{n0}]\pm}\delta_{n,0}~~,
~~F(\mathbf{b},\mathbf{s},\theta^{\pm}|\omega)=A^{[0]\pm}\delta_{n0}~,
\end{equation}
\begin{equation}\label{8-030}
f(\mathbf{b},\mathbf{s},\theta_{n}^{\pm}|\omega)=A_{n}^{[1\pm\delta_{n0}]\pm}~~,
~~F(\mathbf{b},\mathbf{s},\theta^{\pm}|\omega)=A^{[0]\pm}\delta_{n0}~,
\end{equation}
\begin{equation}\label{9-040}
f(\mathbf{b},\mathbf{s},\theta_{n}^{\pm}|\omega)=\mathcal{F}_{n}^{\pm}\delta_{n,0}~~,
~~F(\mathbf{b},\mathbf{s},\theta^{-}|\omega)=\mathfrak{F}^{\pm}\delta_{n0}~,
\end{equation}
\begin{equation}\label{8-050}
f(\mathbf{b},\mathbf{s},\theta_{n}^{-}|\omega)=\mathcal{F}^{}\delta_{n0}~~,
~~F(\mathbf{b},\mathbf{s},\theta^{-}|\omega)=\mathfrak{F}^{\pm}\delta_{n0}~.
\end{equation}
As mentioned previously, only the first two choices are exploitable in the inversions and they give rise to cost functionals that have the following explicit forms:
\begin{equation}\label{8-060}
\kappa_{1}=
\frac{
\|A^{[0]-}-A_{0}^{[0]-}\|^{2}
+
\|A^{[2]+}-A_{0}^{[2]+}\|^{2}
}
{
\|A_{0}^{[0]-}\|^{2}+
\|A_{0}^{[2]+}\|^{2}
}~,
\end{equation}
\begin{equation}\label{8-070}
\kappa_{2}=
\frac{
\sum_{n\in\mathbb{H}^{-}}\|A^{[0]-}\delta_{n0}-A_{n}^{[1-\delta_{n0}]-}\|^{2}
+\sum_{n\in\mathbb{H}^{+}}\|A^{[0]+}\delta_{n0}-A_{n}^{[1+\delta_{n0}]+}\|^{2}
}
{
\sum_{n\in\mathbb{H}^{-}}\|A_{n}^{[1-\delta_{n0}]-}\|^{2}+
\sum_{n\in\mathbb{H}^{+}}\|A_{n}^{[1+\delta_{n0}]+}\|^{2}
}~,
\end{equation}
wherein we have suppressed the dependence of the various terms on $\mathbf{b}$, $\mathbf{s}$, $\mathbf{r}$, $\omega$, $\mathbf{B}$, $\mathbf{S}$, $\mathbf{R}$ in order to make the expressions easier to read.

We wrote earlier that $\mathbf{b}$ is known. This is true, and in fact a necessity, in order to compute $\{A_{n}^{[0]-}\}$, $\{A_{n}^{[2]+}\}$ which is actually the case since what we are trying to find out is how $\tilde{\mathbf{B}}$ is related to $\mathbf{b}$, or, more specifically, how $\tilde{R}^{[1]},~\tilde{C}^{[1]}$ are related to $\rho^{[1]}~,c^{[1]}$ and eventually to the other 'internal' parameters $\rho^{[0]}~,c^{[0]}~;~l=0,1$ and $h,~w,~d$, as well as to the 'external' parameters $A^{i}$, $\theta^{i}$ and $\omega$. If the homogenization scheme, represented by replacing the given  heterogeneous structure by a homogeneous layer, is successful, we would like  $\tilde{R}^{[1]},~\tilde{C}^{[1]}$ to be related to  $\rho^{[1]},~c^{[1]}$ in as simply as possible manner and to be not at all related to the other  (internal or external) parameters.

When an entry of $\mathbf{b}$ is a priori real then we shall assume that the corresponding to-be-retrieved entry of $\mathbf{B}$ is also real.

What appears to be the best procedure is to solve the inverse problem in a 2D search space, but this often leads to instabilities. Herein, we chose to solve for the two real quantities $\tilde{R}^{[1]},~\tilde{C}^{[1]}$ {\it separately}, assuming for each one that the other  takes on  given values. Moreover, for a given set of parameters $\mathbf{b},\mathbf{s},\mathbf{r}$, we assume that the sought-for single entry $B$ (to which corresponds $b$ in $\mathbf{b}$) of $\mathbf{B}$  lies between two values, rather far apart, which constitute the search interval for this quantity, i.e.,
\begin{equation}\label{8-080}
B\in [\underline{B},\overline{B}]~\text{such that~} \underline{B}<<b<<\overline{B}~.
\end{equation}
Even though we assumed (in sect. \ref{bvp1}) sign restrictions on the entries of $\mathbf{b}$, we impose no such restrictions on those of $\mathbf{B}$ (i.e., on those of $\underline{B}$ and $\overline{B}$).

Rather than vary $B$  sequentially, in a haphazard, or in a more intelligent way, we divide the corresponding (1D) search space (actually an interval) into a number of equal subintervals thus corresponding to taking into~consideration~the~discretized values $B_{k}~;~k=1,2,...,N_{B}$ such that $B_{k}=\underline{B}+(k-1)\left(\frac{\overline{B}-\underline{B}}{N_{B}-1}\right)$. The cost functional is then computed for this set  and $\tilde{B}$ is chosen to be the $B_{k}$ corresponding to the smallest (i.e., the global minimum) cost amongst the set of $N_{B}$ costs.
\clearpage
\newpage
\section{Numerical results}
%
\subsection{Anticipated results}
Static homogenization (i.e., mixture models) lead one to believe \cite{si02} that the surrogate layer properties will, at least at low frequencies, follow the laws:
\begin{equation}\label{9-010}
\tilde{R}^{[1]}\approx\tilde{R}_{S}^{[1]}=\rho^{[1]}\alpha_{R}~~,~~\tilde{C}^{[1]}\approx\tilde{C}_{S}^{[1]}= c^{[1]}\alpha_{C}~,
\end{equation}
wherein $\alpha_{R}$ and $\alpha_{C}$ are constants with respect to the frequency $f$. Recall that we assumed herein that $\rho^{[1]}$ and $c^{[1]}$ are likewise constants with respect to $f$ so that $\tilde{R}_{S}^{[1]}$ and $\tilde{C}_{S}^{[1]}$ are constant with respect to $f$.

Another essential aspect of these mixing rules is how $\alpha_{R}$ and $\alpha_{C}$ relate to the structural parameters of the given inhomogeneous medium. The inhomogeneity of a  bicomposite inhomogeneous medium such as ours is usually  characterized by a so-called filling factor $\phi$ which is the proportion (volumetric or areal) of one of the two phases  to the total volume or area of the medium. In our case, $\phi=\Phi(w/d)$ which expresses the fact that $\phi$ is a function of $w/d$ and that we would prefer that it not be a function of the remaining structural parameter $h$.

In a study \cite{wi18}, similar to the present one, we found, by a low-frequency, large filling-factor homogenization procedure, that $\Phi=\phi^{p}$ with $p=0$ (for one of the two constitutive parameters) or $p=1$ (for the other constitutive parameter). This suggests that a suitable low-frequency mixture model for the present problem be of the form
\begin{equation}\label{9-020}
\tilde{R}^{[1]}\approx\tilde{R}_{S}^{[1]}=\rho^{[1]}\alpha_{RS}(\phi)~~,~~\tilde{C}^{[1]}\approx\tilde{C}_{S}^{[1]}= c^{[1]}\alpha_{CS}(\phi)~,
\end{equation}
wherein
\begin{equation}\label{9-030}
\alpha_{RS}(\phi)=\phi^{p_{R}}~~,~~\alpha_{CS}(\phi)=\phi^{p_{C}}~,
\end{equation}
with $p_{R}$ and $p_{C}$ two to-be-determined powers. To carry out this determination, we shall usually assume, as was found in \cite{wi18}, that $p_{C}=0$ and determine the other power by the retrieval procedure at $f=0$. This procedure will be shown to work well for large $\phi$ (whose maximum value is 1) but less-well for small $\phi$.

In general, the retrieval method employed here, whose purpose is to obtain a dynamic homogenization of the given inhomogeneous structure, gives rise to surrogate constitutive parameters which, contrary to $\tilde{R}_{S}^{[1]}$ and $\tilde{C}_{S}^{[1]}$, do depend on the frequency $f$. We shall designate these so-obtained parameters by $\tilde{R}_{D}^{[1]}$ and $\tilde{C}_{D}^{[1]}$ to underline their dispersive nature.
\subsection{Assumed parameters}
In  the following figures it is assumed, unless written otherwise, that:
$\theta^{i}=0^{\circ},~d=0.004~m,~w=0.003~m,~h=0.00475~m,~c^{[0]}=343~ms^{-1},~\rho^{[0]}=1.2~Kgm^{-3},~c^{[1]}=500~ms^{-1},~
\rho^{[1]}=2.4~Kgm^{-3},~c^{[2]}=343~ms^{-1},~\rho^{[2]}=1.2~Kgm^{-3}$.

In each  figure, the abscissa designates the frequency $f$ in $Hz$. The ordinate of the first (uppermost) panel designates a retrieved parameter $\tilde{B}$ of the homogenized layer surrogate. The ordinate of the next (lower) panel designates the  value of the minimal cost function used to choose the retrieved parameter amongst the set of trial parameters. The ordinate of the third panel designates: (i) the rigorously-simulated far-field data $\|A_{0}^{[0]-}\|$ of the given inhomogeneous structure (blue curve), (ii) the reconstructed surrogate layer far-field coefficient $\|A^{[0]-}\|$ obtained by employing $\tilde{R}_{S}^{[1]}$ and $\tilde{C}_{S}^{[1]}$ in the homogeneous layer solution (red curve) and (iii) the reconstructed surrogate layer  far-field coefficient $\|A_{0}^{[0]-}\|$ obtained by employing $\tilde{R}_{D}^{[1]}$ and $\tilde{C}_{D}^{[1]}$ in the layer solution (black curve). The ordinate of the fourth (lowest) panel designates: (i) the rigorously-simulated far-field data $\|A_{0}^{[2]+}\|$ of the given inhomogeneous structure (blue curve), (ii) the reconstructed surrogate layer far-field coefficient $\|A^{[2]+}\|$ obtained by employing $\tilde{R}_{S}^{[1]}$ and $\tilde{C}_{S}^{[1]}$ in the homogeneous layer solution (red curve) and (iii) the reconstructed surrogate layer  far-field coefficient $\|A_{0}^{[2]+}\|$ obtained by employing $\tilde{R}_{D}^{[1]}$ and $\tilde{C}_{D}^{[1]}$ in the layer solution (black curve).
\subsection{Retrieval of the homogenized density $\tilde{R}^{[1]}$}
%
\subsubsection{Comparison of $\tilde{R}^{[1]}$  obtained via $\kappa_{1}$ and $\kappa_{2}$}
\begin{figure}[ht]
\begin{center}
\includegraphics[width=12cm] {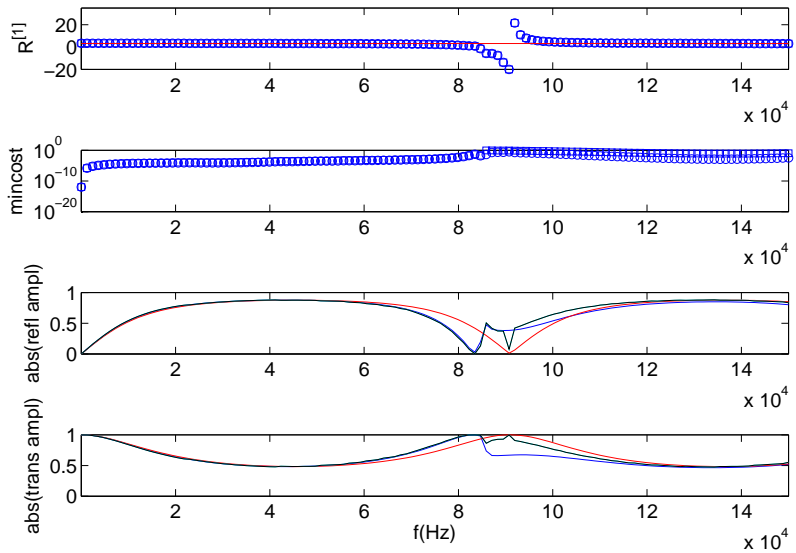}
 \caption{Retrieval of $\tilde{B}=\tilde{R}^{[1]}$. The red line  corresponds to $\tilde{R}_{S}^{[1]}$ assuming $p_{C}=0$ and $p_{R}=-1$. The circles correspond to the retrieval of $\tilde{R}_{D}^{[1]}$ via cost functional $\kappa_{1}$ whereas the diamonds correspond to the retrieval of $\tilde{R}_{S}^{[1]}$ via cost functional $\kappa_{2}$. The blue curves in the third and fourth panels correspond to the rigorously-simulated coefficients of the far-field data, the red curves to the reconstructed far-field coefficients obtained with $\tilde{R}_{S}^{[1]}=\rho^{[1]}\phi^{-1},~\tilde{C}_{S}^{[1]}=c^{[1]}$ and the black curves to the reconstructed far-field coefficients obtained with $\tilde{R}_{D}^{[1]}$. Case $h=0.00275~m$.}
  \label{k1k2-10}
  \end{center}
\end{figure}

The object here, illustrated in fig. \ref{k1k2-10}, is to find out whether there exists a difference between the retrievals of $\tilde{B}=\tilde{R}^{[1]}$ obtained by minimizing $\kappa_{2}$ and those obtained by minimizing $\kappa_{1}$.

We note that the retrievals are (graphically) identical  (even though there is some difference in the minima of the cost functions) be they obtained via $\kappa_{1}$ or $\kappa_{2}$. Moreover,
 the reconstructed layer coefficients $\|A^{[0]-}\|$ and $\|A^{[2]+}\|$ using the dispersive retrieved parameter  obtained by minimizing $\kappa_{2}$ are (graphically (they overlap the black curves)) identical to those obtained by minimizing  $\kappa_{1}$, so that from now  the retrievals of $\tilde{B}=\tilde{R}^{[1]}$ will be obtained by minimizing $\kappa_{1}$.
\subsubsection{Retrieval of $\tilde{R}_{D}^{[1]}$  for various choices of $C^{[1]}$}
\begin{figure}[ht]
\begin{center}
\includegraphics[width=12cm] {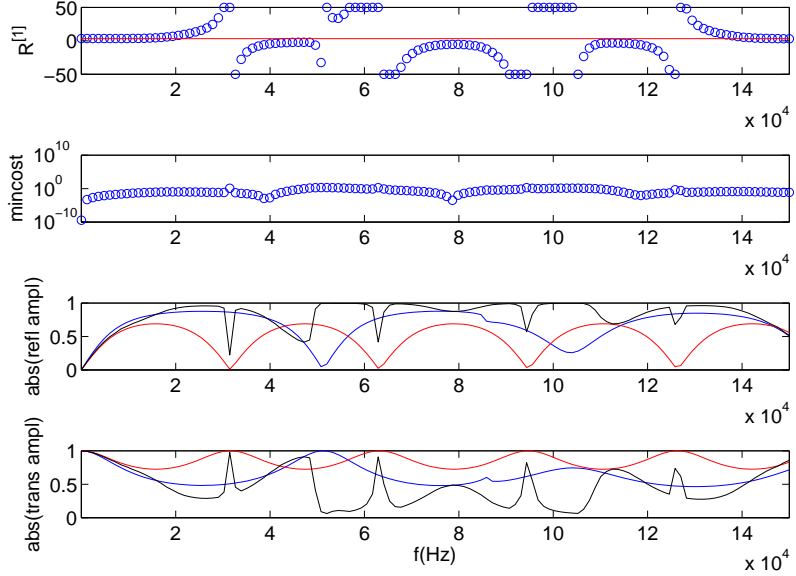}
 \caption{Retrieval of $\tilde{B}=\tilde{R}^{[1]}$. The red line  corresponds to $\tilde{R}_{S}^{[1]}$ assuming $p_{C}=0$ and $p_{R}=-1$. The circles correspond to the retrieval of $\tilde{R}_{D}^{[1]}$ via cost functional $\kappa_{1}$. The blue curves in the third and fourth panels correspond to the rigorously-simulated coefficients of the far-field data, the red curves to the reconstructed far-field coefficients obtained with $\tilde{R}_{S}^{[1]}=\rho^{[1]}\phi^{-1},~\tilde{C}_{S}^{[1]}=c^{[1]}$ and the black curves to the reconstructed far-field coefficients obtained with $\tilde{R}_{D}^{[1]}$. The filling factor is $\phi=3/4$. $h=0.00475~m$. Choice $C^{[1]}=300~ms^{-1}$ to retrieve $\tilde{R}_{D}^{[1]}$.}
  \label{C1onR1-10}
  \end{center}
\end{figure}

\begin{figure}[ptb]
\begin{center}
\includegraphics[width=12cm] {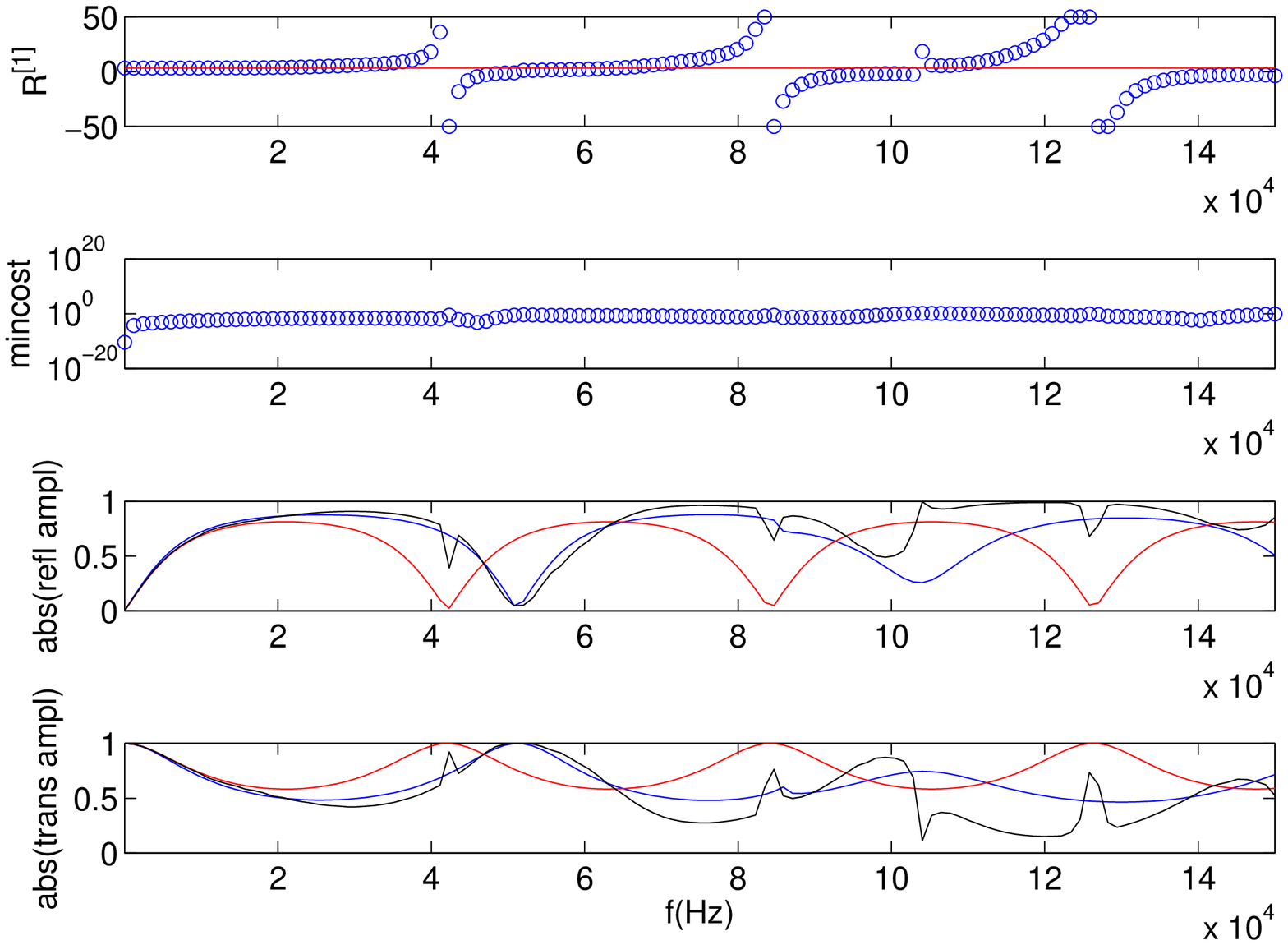}
 \caption{Same as fig. \ref{C1onR1-10} for the choice $C^{[1]}=400~ms^{-1}$ to retrieve $\tilde{R}_{D}^{[1]}$.}
  \label{C1onR1-20}
  \end{center}
\end{figure}

\begin{figure}[ptb]
\begin{center}
\includegraphics[width=12cm] {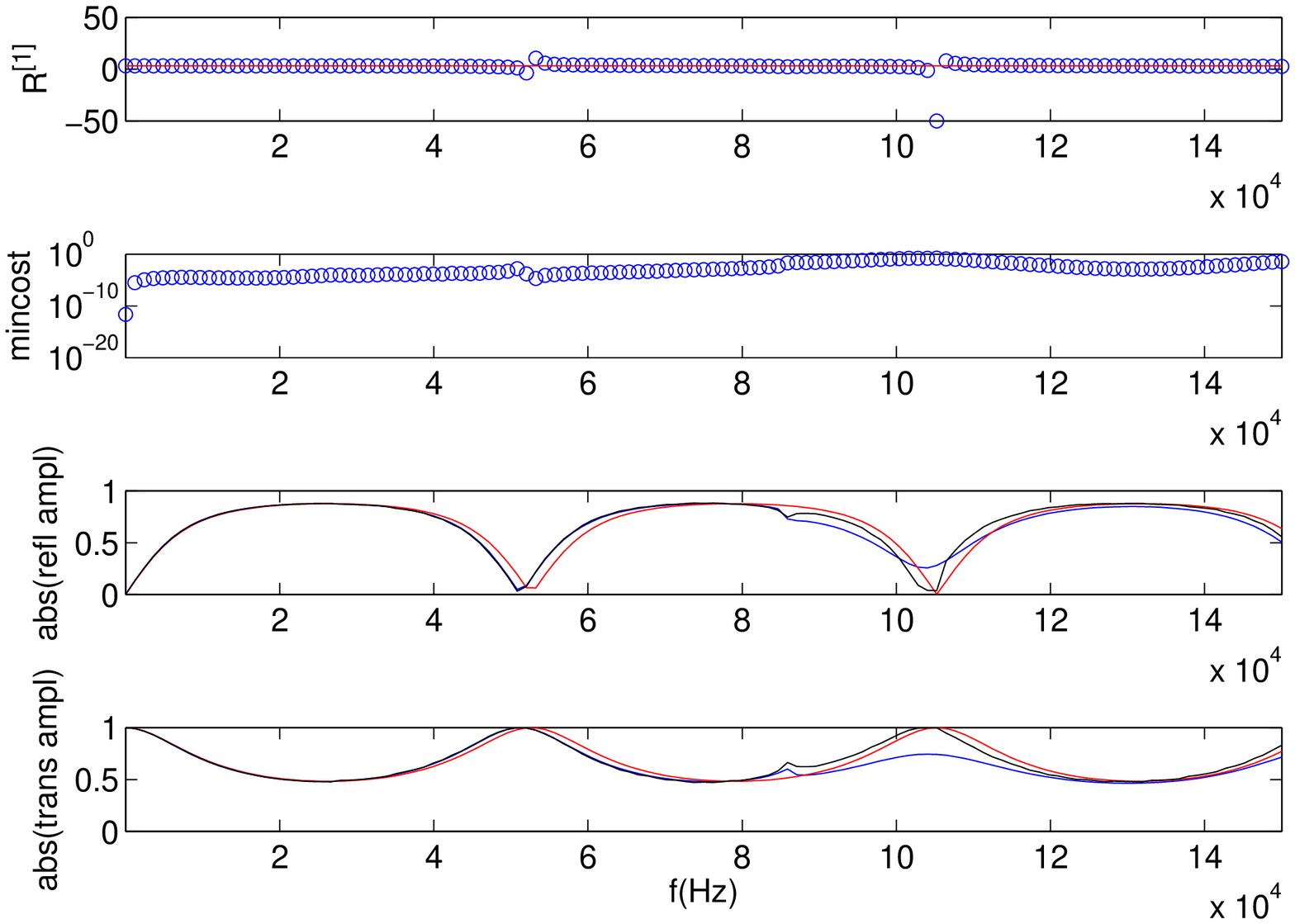}
 \caption{Same as fig. \ref{C1onR1-10} for the choice $C^{[1]}=c^{[1]}=500~ms^{-1}$ to retrieve $\tilde{R}_{D}^{[1]}$.}
  \label{C1onR1-30}
  \end{center}
\end{figure}

\begin{figure}[ptb]
\begin{center}
\includegraphics[width=12cm] {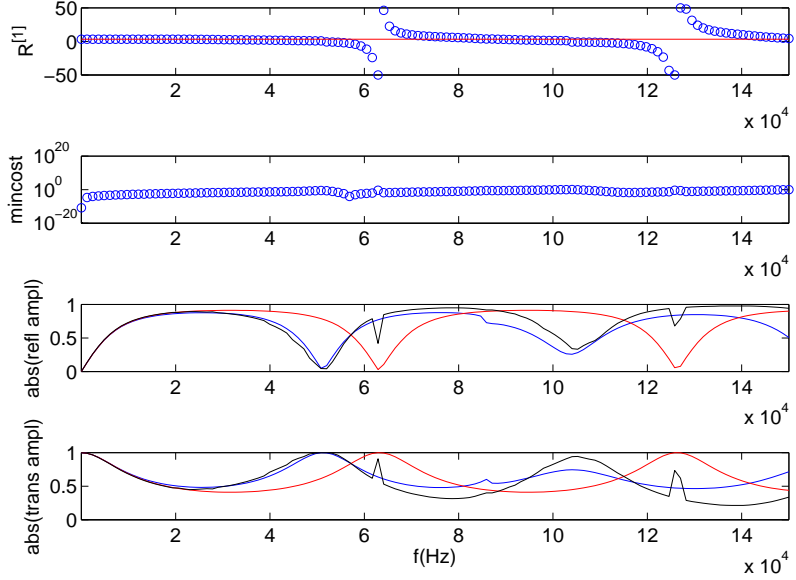}
 \caption{Same as fig. \ref{C1onR1-10} for the choice $C^{[1]}=600~ms^{-1}$ to retrieve $\tilde{R}_{D}^{[1]}$.}
  \label{C1onR1-40}
  \end{center}
\end{figure}

As specified earlier, in order to obtain $\tilde{R}_{D}^{[1]}$ we must specify  $\tilde{C}^{[1]}$. The impact on $\tilde{R}_{D}^{[1]}$ of various choices of $C^{[1]}$ is depicted in figs.  \ref{C1onR1-10}-\ref{C1onR1-40} wherein it is clearly seen that the $\tilde{R}_{D}^{[1]}$ with the least perturbative features as a function of frequency and the best agreement with reconstructed coefficients corresponds to the choice $\tilde{C}^{[1]}=c^{[1]}=500~ms^{-1}$.
\newpage
 Moreover, with this choice, the low-frequency behavior of $\tilde{R}_{D}^{[1]}$ is consistent with that of a horizontal line whose height is $3.2~Kgm^{-3}=2.4\times 4/3~~Kgm^{-3}$ or $p_{R}=-1$. This explains why the horizontal red lines in these four figures are situated where they are.

 These and other results not shown here seem to indicate that the best choice for obtaining $\tilde{R}_{D}^{[1]}$ is  $C^{[1]}=c^{[1]}$. This choice will be made in all the subsequent retrievals of $\tilde{R}^{[1]}$ unless indicated otherwise.

 A last important remark concerning fig.  \ref{C1onR1-30}: the reconstructed surrogate layer far-field response obtained with $\tilde{R}_{S}^{[1]}$ is in good agreement with the simulated far-field data over a rather broad range of low frequencies.
\clearpage
\newpage
\subsubsection{Retrieval of $\tilde{R}_{D}^{[1]}$ for various  layer thicknesses $h$}
\begin{figure}[ht]
\begin{center}
\includegraphics[width=12cm] {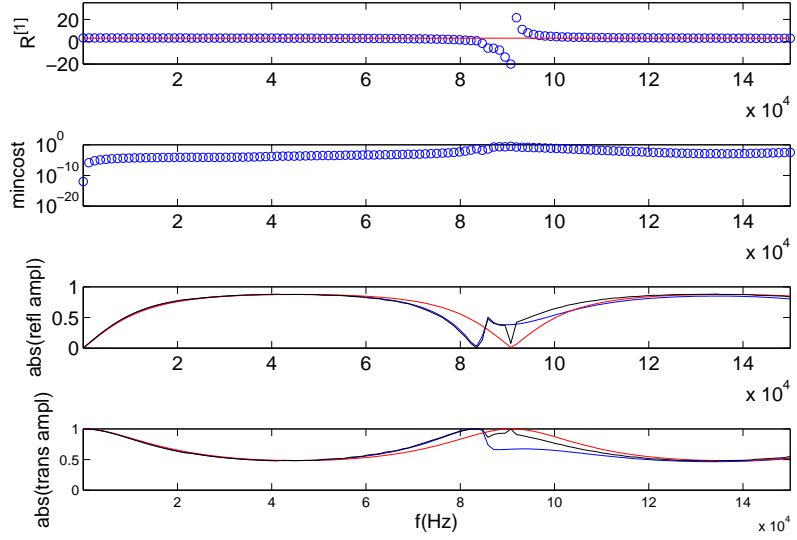}
 \caption{Retrieval of $\tilde{B}=\tilde{R}^{[1]}$. The red line corresponds to $\tilde{R}_{S}^{[1]}$ assuming $p_{C}=0$ and $p_{R}=-1$. The circles correspond to the retrieval of $\tilde{R}_{D}^{[1]}$ via cost functional $\kappa_{1}$. The blue curves in the third and fourth panels correspond to the rigorously-simulated coefficients of the far-field data, the red curves to the reconstructed far-field coefficients obtained with $\tilde{R}_{S}^{[1]}$ and the black curves to the reconstructed far-field coefficients obtained with $\tilde{R}_{D}^{[1]}$. The filling factor is $\phi=3/4$. $C^{[1]}=c^{[1]}$. Case $h=0.00275~m$.}
  \label{honR1-10}
  \end{center}
\end{figure}

\begin{figure}[ptb]
\begin{center}
\includegraphics[width=12cm] {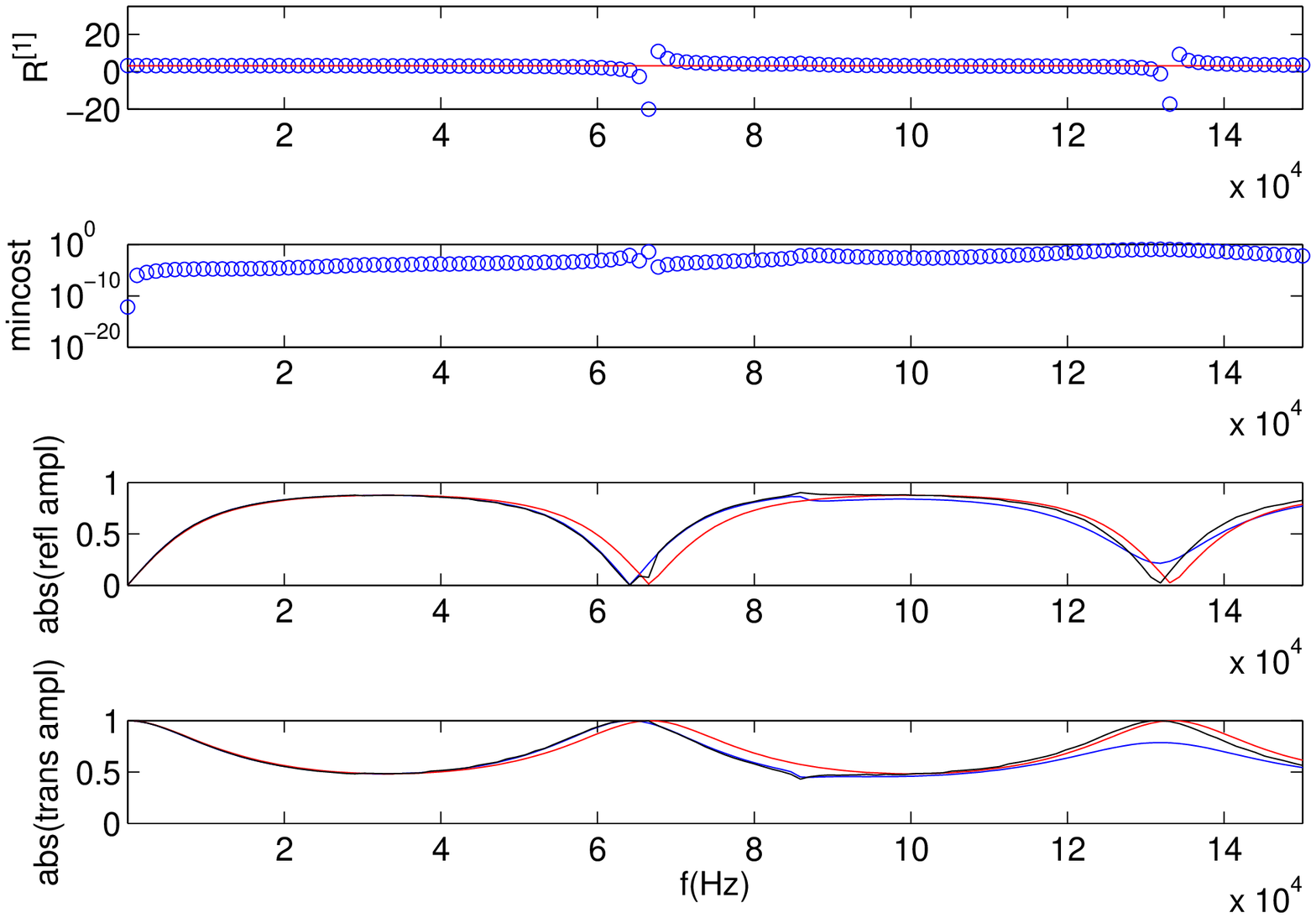}
 \caption{Same as fig. \ref{honR1-10} for the case $h=0.00375~m$.}
  \label{honR1-20}
  \end{center}
\end{figure}

\begin{figure}[ptb]
\begin{center}
\includegraphics[width=12cm] {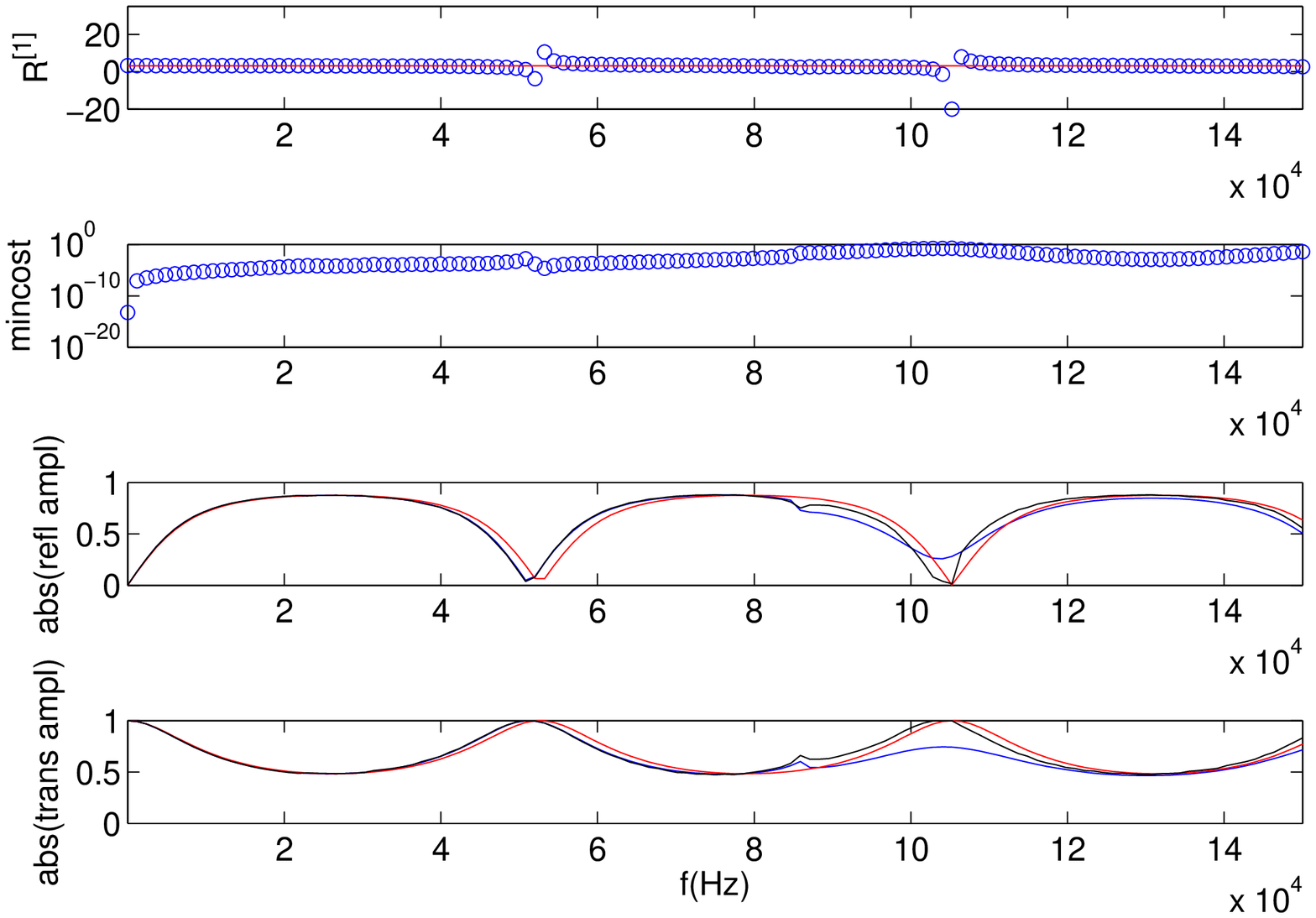}
 \caption{Same as fig. \ref{honR1-10} for the case  $h=0.00475~m$.}
  \label{honR1-30}
  \end{center}
\end{figure}

As written earlier, we would like the homogenized constitutive parameters to not depend on the given structure thickness=surrogate layer thickness=$h$. Figs.  \ref{honR1-10}-\ref{honR1-30} tell us how $\tilde{R}_{D}^{[1]}$ varies with $h$.
\newpage
Although the variation of $h$ produces variations in the far-field responses (see two lower panels in the figures) this has practically no effect on $\tilde{R}_{D}^{[1]}$ except in the close neighborhoods of the frequencies at which the far-field reflected response is nil, a situation that results in the inversion procedure giving rise to a resonant type of feature. The latter enables a rather good agreement between the far-field data and reconstructed far-field response which is much better than that obtained via $\tilde{R}_{S}^{[1]}$. However, the reconstructed surrogate layer far-field response obtained with $\tilde{R}_{S}^{[1]}$ is in good agreement with the simulated far-field data over a rather broad range of low frequencies.

Last  but not least we note that  $R_{D}^{[1]}=R_{S}^{[1]}=3.2~Kgm^{-3}$ (corresponding to $p_{R}=-1$ over a broad range of frequencies.
\subsubsection{Retrieval of $\tilde{R}_{D}^{[1]}$ for increasing $\theta^{i}$}
\begin{figure}[ht]
\begin{center}
\includegraphics[width=12cm] {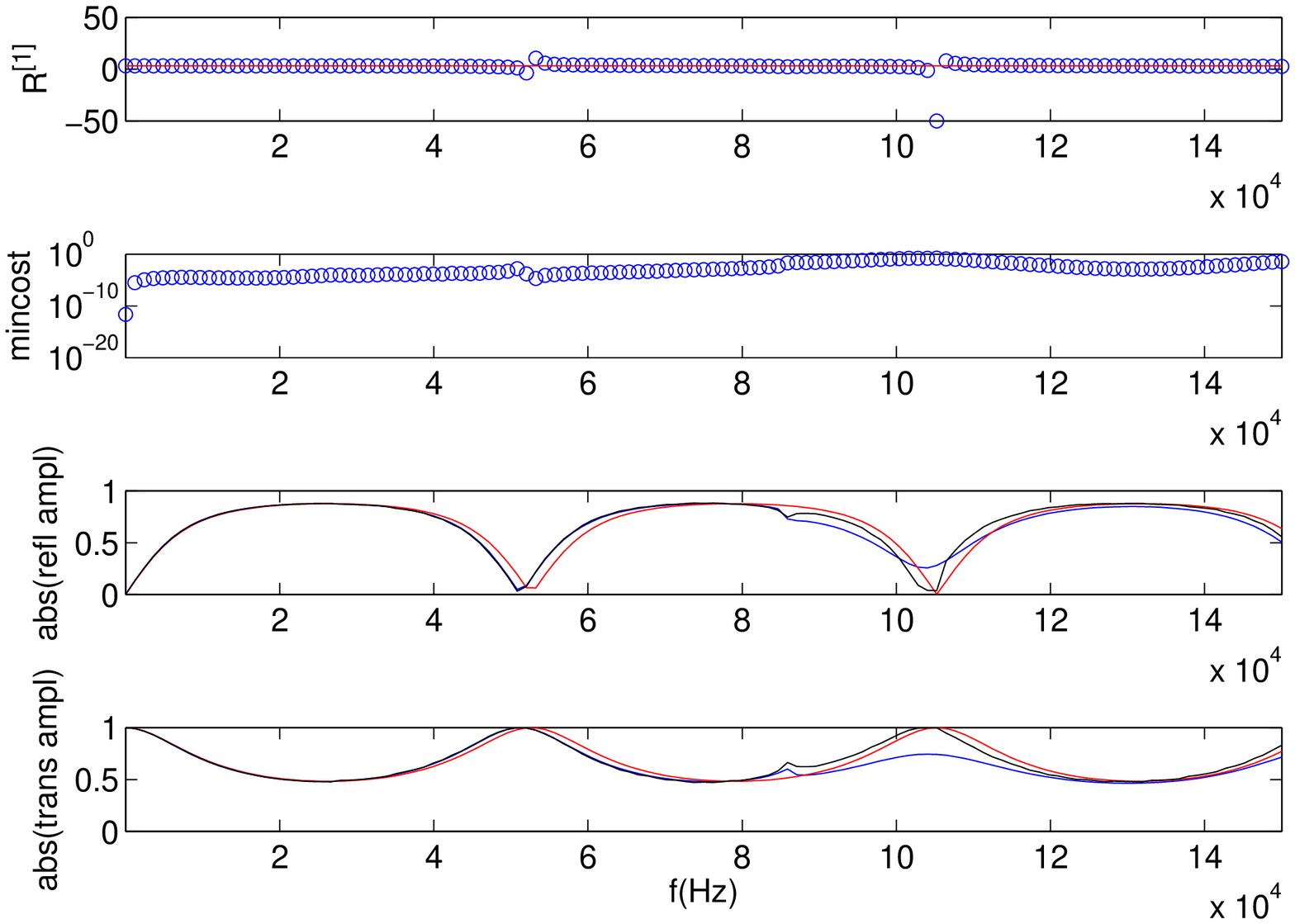}
 \caption{Retrieval of $\tilde{B}=\tilde{R}^{[1]}$. The red line corresponds to $\tilde{R}_{S}^{[1]}$ assuming $p_{C}=0$ and $p_{R}=-1$. The circles correspond to the retrieval of $\tilde{R}_{D}^{[1]}$ via cost functional $\kappa_{1}$. The blue curves in the third and fourth panels correspond to the rigorously-simulated coefficients of the far-field data, the red curves to the reconstructed far-field coefficients obtained with $\tilde{R}_{S}^{[1]}$ and the black curves to the reconstructed far-field coefficients obtained with $\tilde{R}_{D}^{[1]}$.  $C^{[1]}=c^{[1]}$. $w=0.003~m$. $h=0.00475~m$. Case $\theta^{i}=0^{\circ}$.}
  \label{thonR1-10}
  \end{center}
\end{figure}

\begin{figure}[ptb]
\begin{center}
\includegraphics[width=12cm] {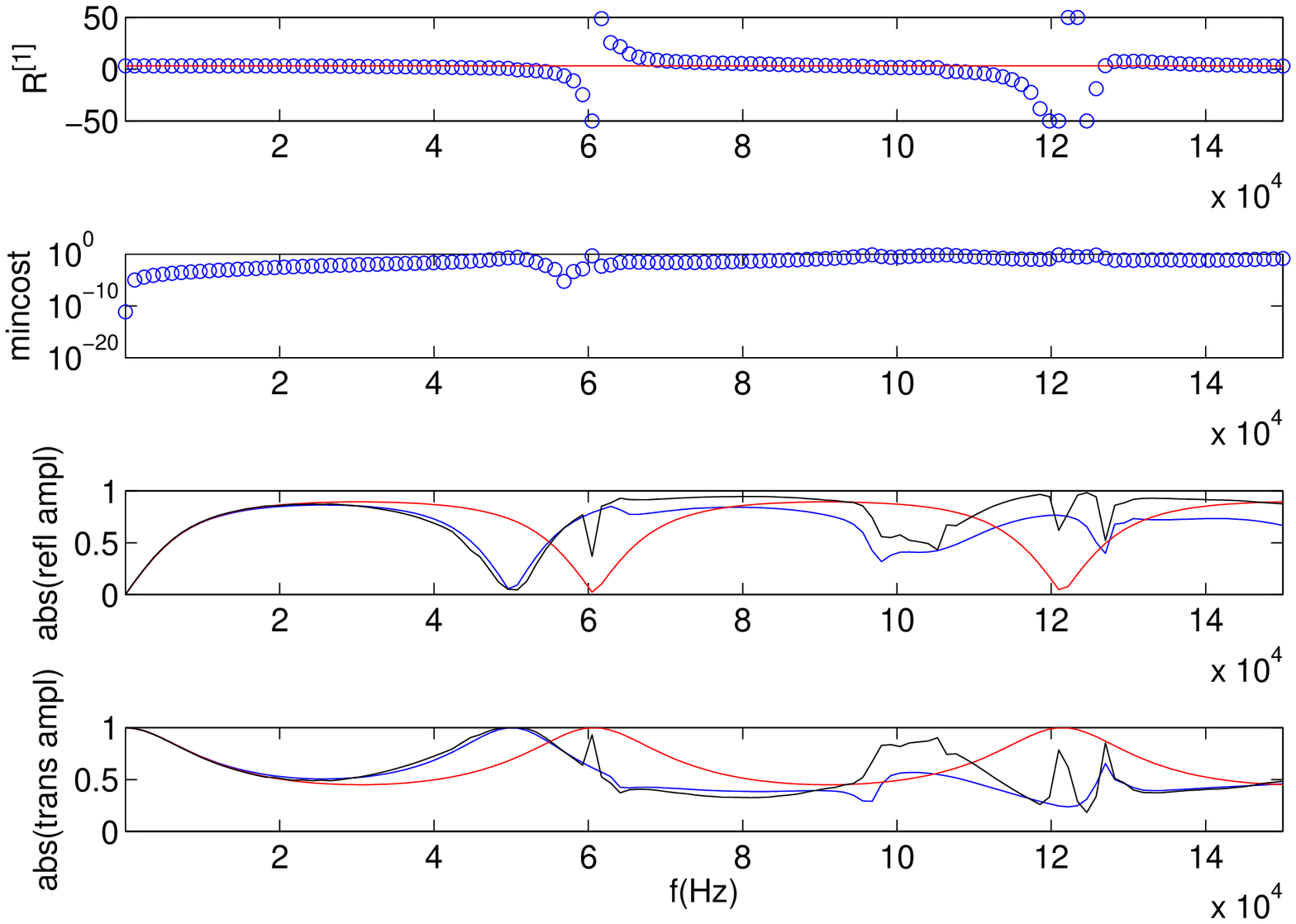}
 \caption{Same as fig. \ref{thonR1-10} for the case $\theta^{i}=20^{\circ}$.}
  \label{thonR1-20}
  \end{center}
\end{figure}

\begin{figure}[ptb]
\begin{center}
\includegraphics[width=12cm] {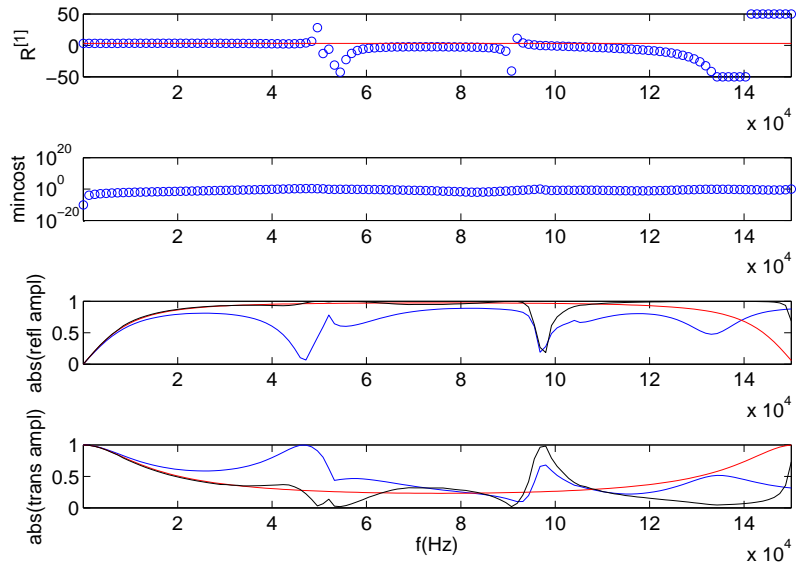}
 \caption{Same as fig. \ref{thonR1-10} for the case $\theta^{i}=40^{\circ}$.}
  \label{thonR1-30}
  \end{center}
\end{figure}

As indicated earlier, we would like the homogenized constitutive parameters to not depend on the incident angle $\theta^{i}$.  Figs.  \ref{thonR1-10}-\ref{thonR1-30} tell us how $\tilde{R}_{D}^{[1]}$ varies with $\theta^{i}$.
\newpage
Although the increase of $\theta^{i}$ produces  increasing discrepancies between the far-field data and the reconstructed  far-field responses (see two lower panels in the figures) this is not very noticeable in the variations of $R_{D}^{[1]}$ with this external parameter except for the anomalous features that become more pronounced with increasing $\theta^{i}$. A remarkable feature of these figures is that they show how $R_{D}^{[1]}$ results in much better agreement than does $R_{S}^{[1]}$ between the far-field data and the reconstructed far-field response, which fact points to the "curative virtue" of dispersion introduced by the retrieval method. However, the curative power of this  'induced dispersion' \cite{wi16a,wi16b,wi16c} has its limits as seen in fig.  \ref{thonR1-30}. In fact, for the $40^{\circ}$ angle of incidence, the homogenized layer, be it dispersive or non-dispersive, responds very differently from the given inhomogeneous structure except at very low frequencies.

Thus, it is not possible to conclude that our homogenized layer responds independently of $\theta^{i}$, this being due to the fact that our dynamic homogenization recipe does not take into account the external parameter $\theta^{i}$.
\subsubsection{Retrieval of ${R}_{D}^{[1]}$ for constant $d$ and increasing $w$}
\begin{figure}[ht]
\begin{center}
\includegraphics[width=12cm] {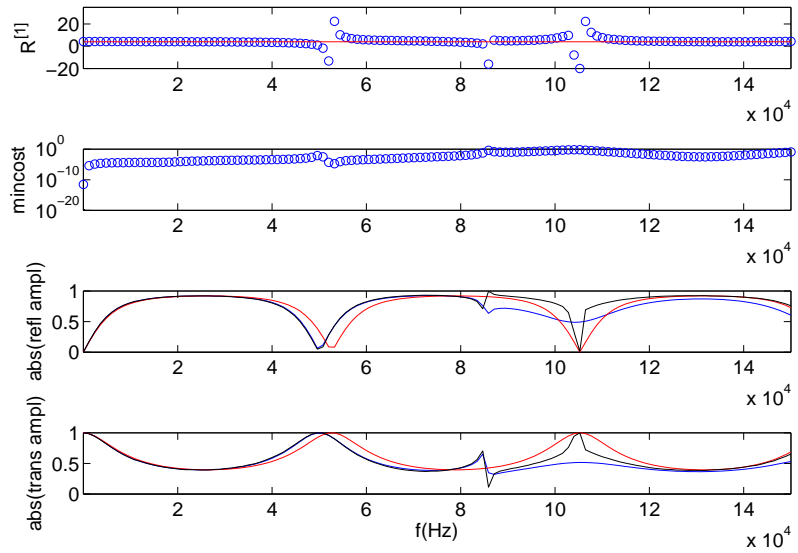}
 \caption{Retrieval of $\tilde{B}=\tilde{R}^{[1]}$. The red line corresponds to $\tilde{R}_{S}^{[1]}$ assuming $p_{C}=0$ and $p_{R}=-1$. The circles correspond to the retrieval of $\tilde{R}_{D}^{[1]}$ via cost functional $\kappa_{1}$. The blue curves in the third and fourth panels correspond to the rigorously-simulated coefficients of the far-field data, the red curves to the reconstructed far-field coefficients obtained with $\tilde{R}_{S}^{[1]}$ and the black curves to the reconstructed far-field coefficients obtained with $\tilde{R}_{D}^{[1]}$.  $h=0.00475~m$. $\theta^{i}=0^{\circ}$. $C^{[1]}=c^{[1]}$. Case $w=0.0024~m$ ($\phi=0.6$).}
  \label{wonR1-10}
  \end{center}
\end{figure}

\begin{figure}[ptb]
\begin{center}
\includegraphics[width=12cm] {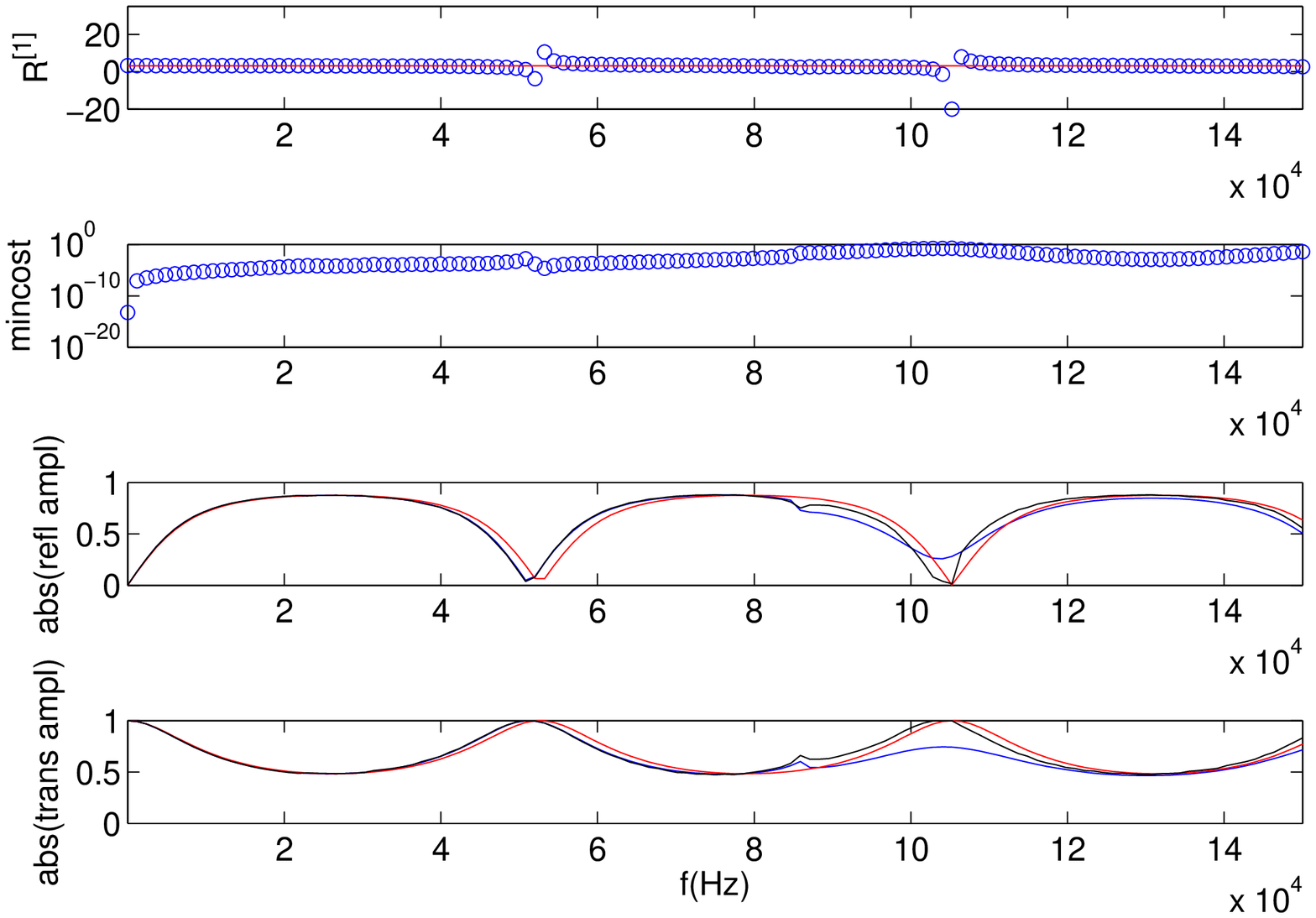}
 \caption{Same as fig. \ref{wonR1-10} for the case $w=0.0030~m$ ($\phi=0.75$).}
  \label{wonR1-20}
  \end{center}
\end{figure}

\begin{figure}[ptb]
\begin{center}
\includegraphics[width=12cm] {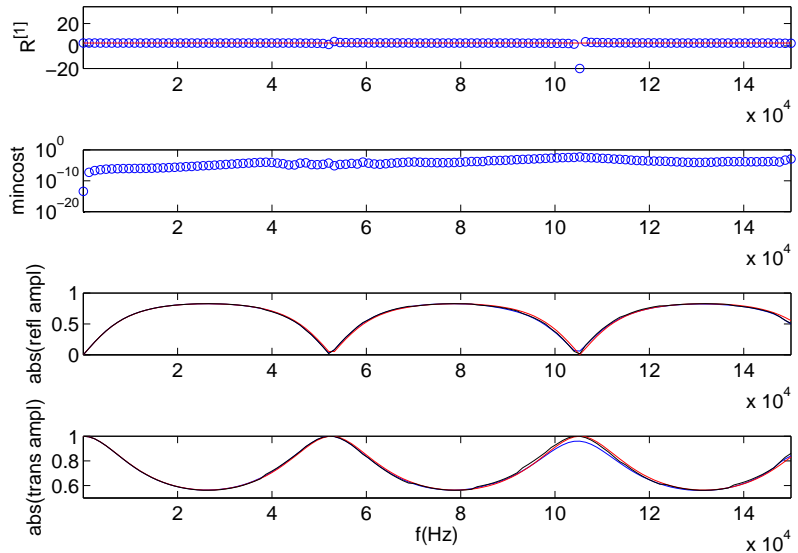}
 \caption{Same as fig. \ref{wonR1-10} for the case $w=0.0036~m$ ($\phi=0.9$).}
  \label{wonR1-30}
  \end{center}
\end{figure}

There exist good reasons to believe \cite{wi18} that our homogenization scheme works less well for smaller filling factors $\phi$. To address this issue, we kept $d$ constant ($=0.004~m$ as usual) and varied $w$, which gave rise to the results in figs.  \ref{wonR1-10}-\ref{wonR1-30}.
\clearpage
\newpage
This is substantiated by the results in these figures wherein we note once again the curative effect of taking into account the dispersion of the retrieved density for the reconstructed far-field response, especially at low $\phi$. We also note that the dispersive layer model,  contrary to the static layer model, is able to account for the existence (near $f=86~KHz$) of Wood anomalies \cite{ma16} which occur near the frequencies at which a plane wave in either the reflected or tansmitted field of the grating changes its nature from homogeneous (body wave) to inhomogeneous (surface wave).
\subsection{Retrieval of the homogenized velocity  $\tilde{C}^{[1]}$}
%
\subsubsection{Comparison of  $\tilde{C}^{[1]}$ obtained via $\kappa_{1}$ and $\kappa_{2}$}
\begin{figure}[ht]
\begin{center}
\includegraphics[width=12cm] {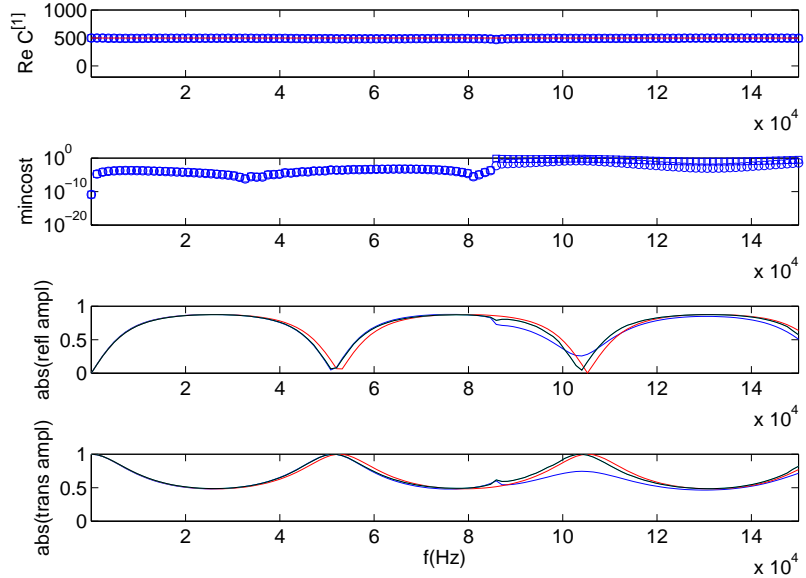}
 \caption{Retrieval of $\tilde{B}=\tilde{C}^{[1]}$. The red line  corresponds to $\tilde{C}_{S}^{[1]}$ assuming $p_{C}=0$ and $p_{R}=-1$. The circles correspond to the retrieval of $\tilde{C}_{D}^{[1]}$ via cost functional $\kappa_{1}$ whereas the diamonds correspond to the retrieval of $\tilde{C}_{S}^{[1]}$ via cost functional $\kappa_{2}$. The blue curves in the third and fourth panels correspond to the rigorously-simulated coefficients of the far-field data, the red curves to the reconstructed far-field coefficients  obtained with $\tilde{R}_{S}^{[1]}=\rho^{[1]}\phi^{-1},~\tilde{C}_{S}^{[1]}=c^{[1]}$  and the black curves to the reconstructed far-field coefficients obtained with $\tilde{C}_{D}^{[1]}$. The filling factor is $\phi=3/4$. Case $h=0.00475~m$.}
  \label{k1k2-20}
  \end{center}
\end{figure}

The object here, illustrated in fig. \ref{k1k2-20}, is to find out whether there exists a difference between the retrievals of $\tilde{B}=\tilde{C}^{[1]}$ obtained by minimizing $\kappa_{2}$ and those obtained by minimizing $\kappa_{1}$.
We note that the retrievals are (graphically) identical  (even though there is some difference in the minima of the cost functions) be they obtained via $\kappa_{1}$ or $\kappa_{2}$. Moreover,
 the reconstructed layer coefficients $\|A_{0}^{[0]-}\|$ and $\|A_{2}^{[0]+}\|$ using the dispersive retrieved parameter  obtained by minimizing $\kappa_{2}$ are (graphically (they overlap the black curves)) identical to those obtained by minimizing  $\kappa_{1}$, so that from now  the retrievals of $\tilde{B}=\tilde{C}^{[1]}$ will be obtained by minimizing $\kappa_{1}$.
\subsubsection{Retrieval of $\tilde{C}^{[1]}_{D}$  for various choices of $R^{[1]}$}
\begin{figure}[ht]
\begin{center}
\includegraphics[width=12cm] {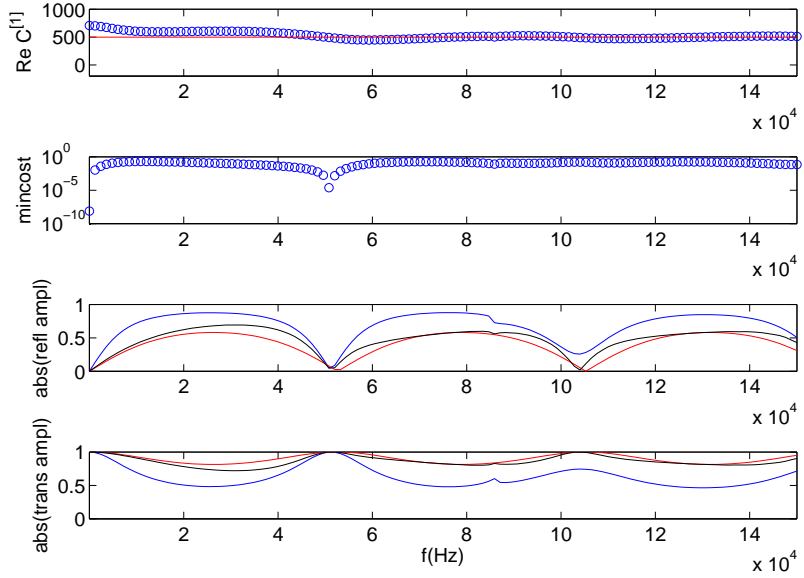}
 \caption{Retrieval of $\tilde{B}=\tilde{C}^{[1]}$. The red line corresponds to $\tilde{C}_{S}^{[1]}$ assuming $p_{C}=0$ and $p_{R}=-1$. The circles correspond to the retrieval of $\tilde{C}_{D}^{[1]}$ via cost functional $\kappa_{1}$. The blue curves in the third and fourth panels correspond to the rigorously-simulated coefficients of the far-field data, the red curves to the reconstructed far-field coefficients obtained with $\tilde{R}_{S}^{[1]}=\rho^{[1]}\phi^{-1},~\tilde{C}_{S}^{[1]}=c^{[1]}$  and the black curves to the reconstructed far-field coefficients obtained with $\tilde{C}_{D}^{[1]}$. The filling factor is $\phi=3/4$. $h=0.00475~m$. Choice $R^{[1]}=1.6~Kgm^{-3}$.}
  \label{R1onC1-10}
  \end{center}
\end{figure}

\begin{figure}[ptb]
\begin{center}
\includegraphics[width=12cm] {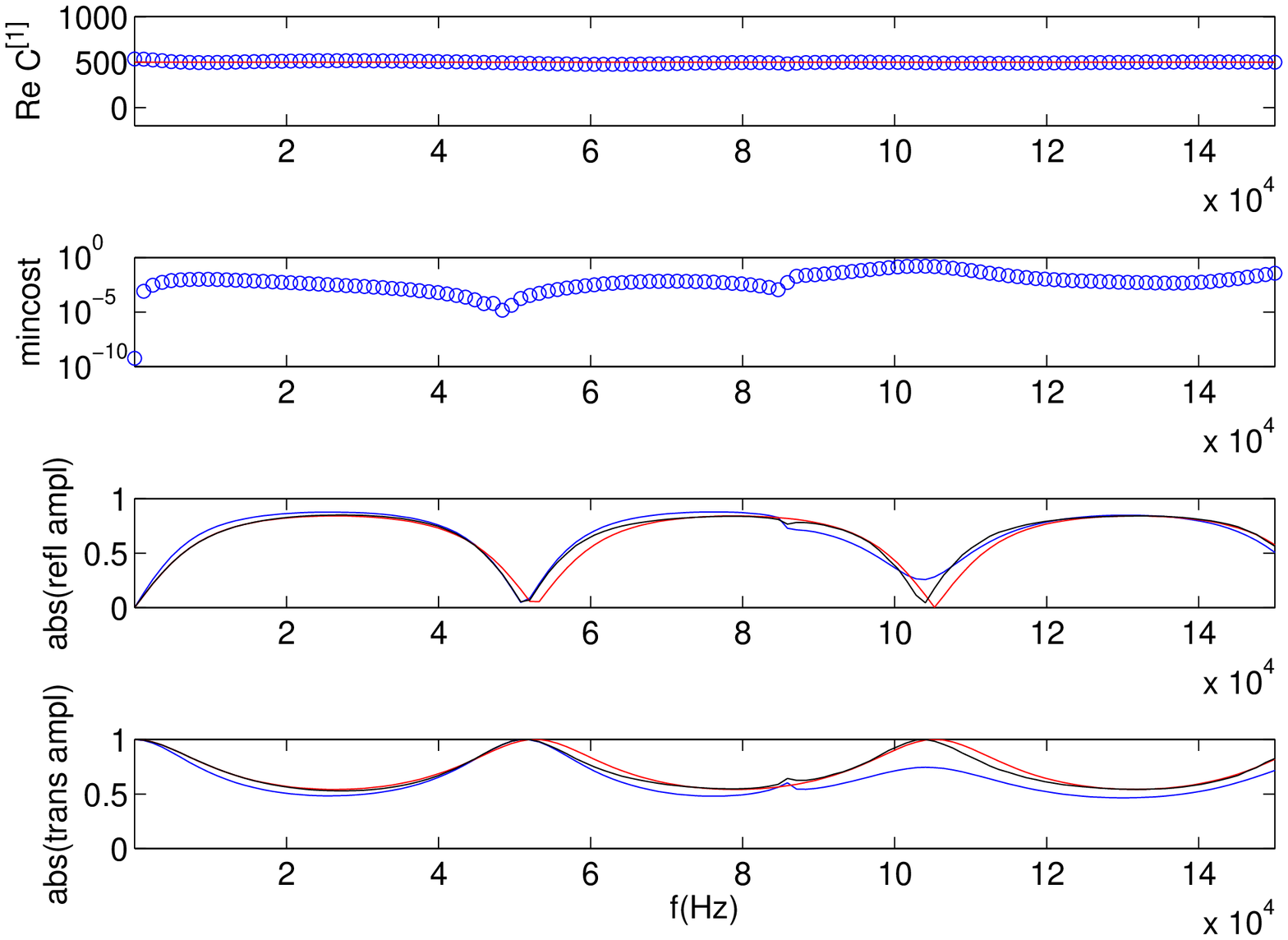}
 \caption{Same as fig. \ref{R1onC1-10} for the choice $R^{[1]}=2.8~Kgm^{-3}$.}
  \label{R1onC1-20}
  \end{center}
\end{figure}

\begin{figure}[ptb]
\begin{center}
\includegraphics[width=12cm] {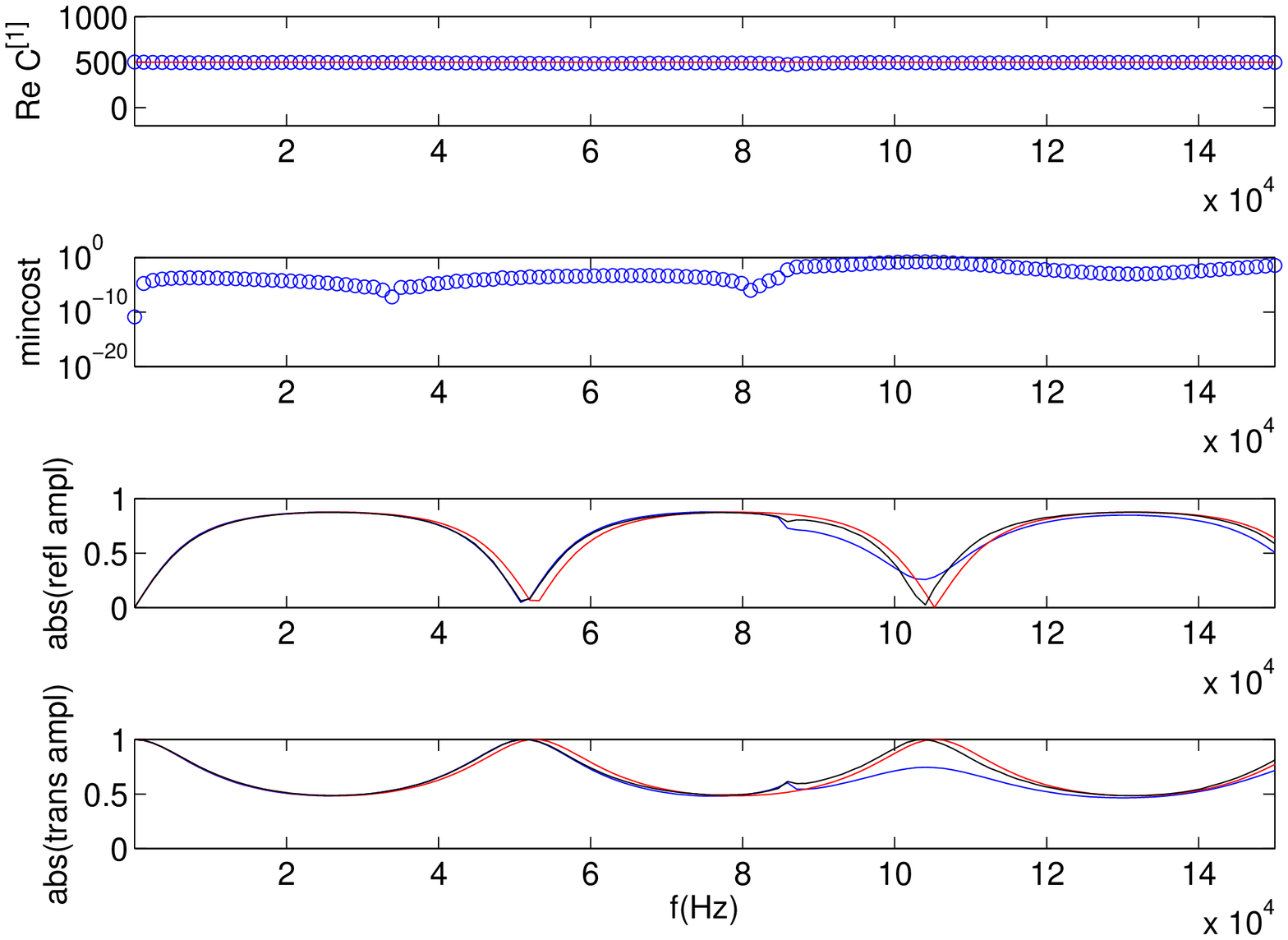}
 \caption{Same as fig. \ref{R1onC1-10} for the choice $R^{[1]}=3.2~Kgm^{-3}$.}
  \label{R1onC1-30}
  \end{center}
\end{figure}

\begin{figure}[ptb]
\begin{center}
\includegraphics[width=12cm] {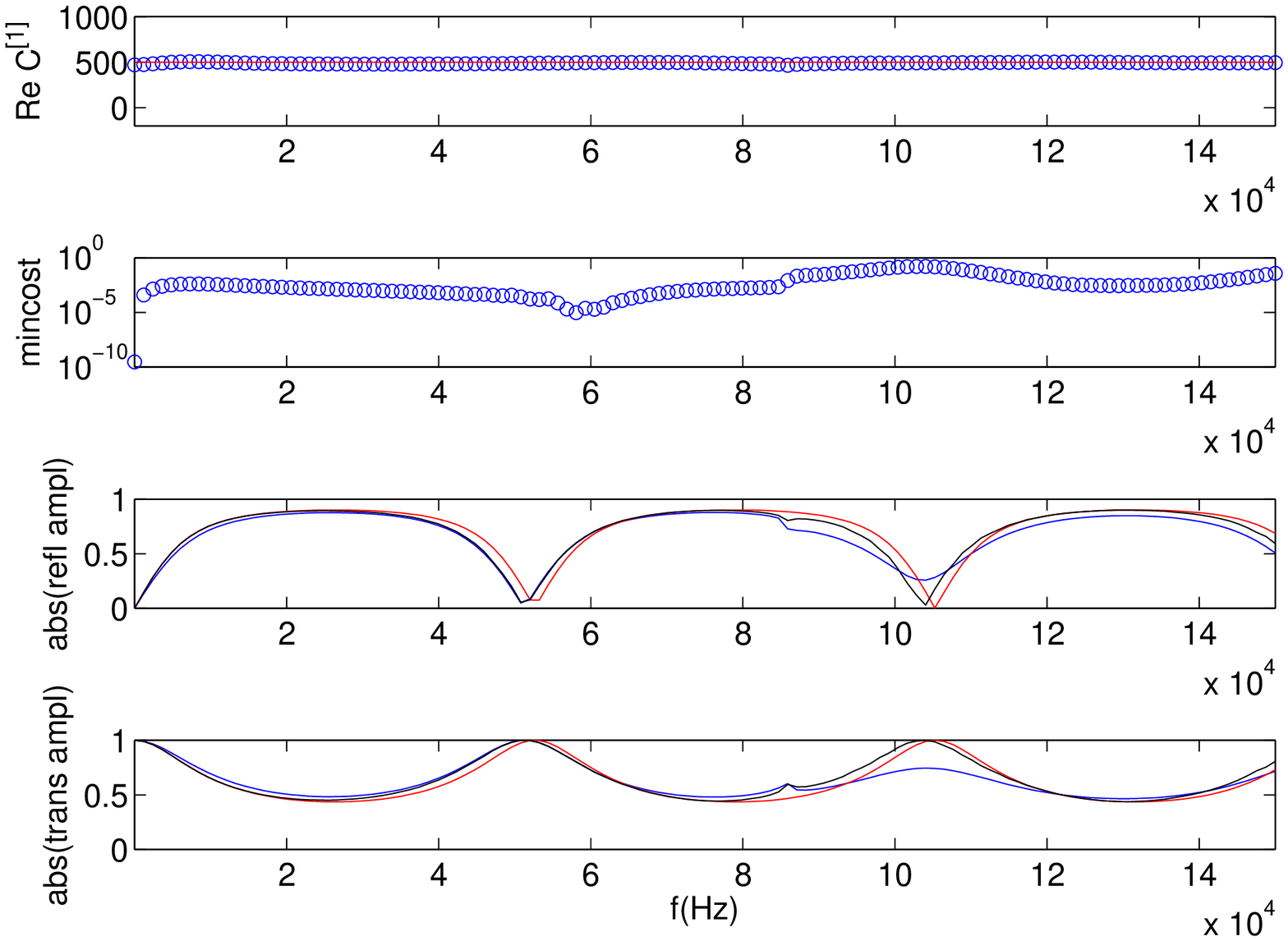}
 \caption{Same as fig. \ref{R1onC1-10} for the choice $R^{[1]}=3.6~Kgm^{-3}$.}
  \label{R1onC1-40}
  \end{center}
\end{figure}

\begin{figure}[ptb]
\begin{center}
\includegraphics[width=12cm] {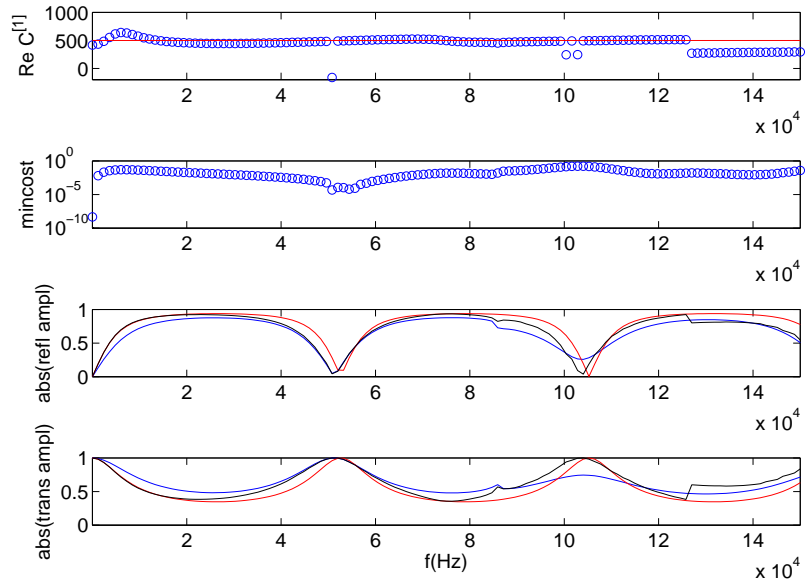}
 \caption{Same as fig. \ref{R1onC1-10} for the choice $R^{[1]}=4.6~Kgm^{-3}$.}
  \label{R1onC1-50}
  \end{center}
\end{figure}

As specified earlier, in order to obtain $\tilde{C}_{D}^{[1]}$ we must specify  $\tilde{R}^{[1]}$. The impact on $\tilde{C}_{D}^{[1]}$ of various choices of $R^{[1]}$ is depicted in figs.  \ref{R1onC1-10}-\ref{R1onC1-50} wherein it is clearly seen that the $\tilde{C}_{D}^{[1]}$ with the least perturbative features as a function of frequency and the best agreement with reconstructed coefficients corresponds to the choice $\tilde{R}^{[1]}=\rho^{[1]}\phi^{-1}=3.2~Kgm^{-3}$.
\newpage
 Moreover, with this choice, the low-frequency behavior of $\tilde{C}_{D}^{[1]}$ is consistent with that of a horizontal line whose height is $500~m$ or $p_{C}=0$. This explains why the horizontal red lines in these four figures are situated where they are.

 These and other results not shown here seem to indicate that the best choice for obtaining $\tilde{C}^{[1]}$ is  $R^{[1]}=\rho^{[1]}\phi^{-1}$. This choice will be made in all the subsequent retrievals of $\tilde{C}^{[1]}$ unless indicated otherwise.

 A last important remark concerning fig.  \ref{R1onC1-30}: the reconstructed surrogate layer far-field response obtained with $\tilde{C}_{S}^{[1]}$ is in good agreement with the simulated far-field data over a rather broad range of low frequencies.
\subsubsection{Retrieval of $\tilde{C}^{[1]}_{D}$  for  increasing $h$}
\begin{figure}[ht]
\begin{center}
\includegraphics[width=12cm] {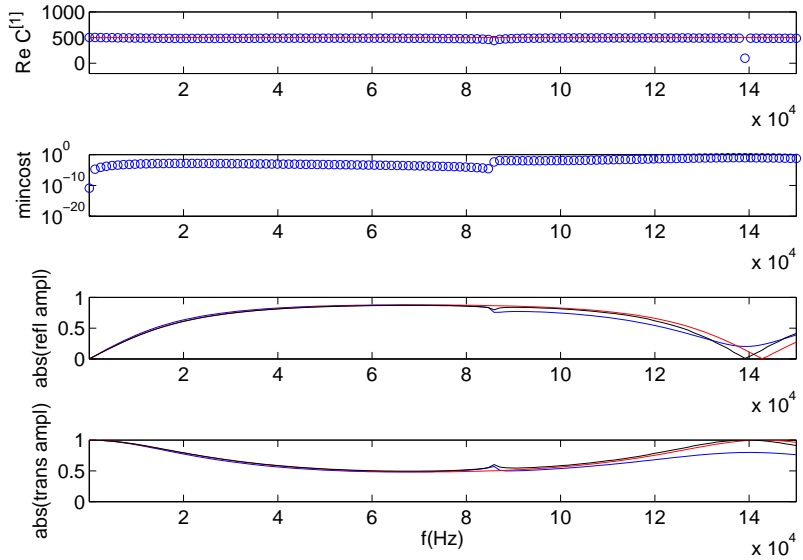}
 \caption{Retrieval of $\tilde{B}=\tilde{C}^{[1]}$. The red line corresponds to $\tilde{C}_{S}^{[1]}$ assuming $p_{C}=0$ and $p_{R}=-1$. The circles correspond to the retrieval of $\tilde{C}_{D}^{[1]}$ via cost functional $\kappa_{1}$. The blue curves in the third and fourth panels correspond to the rigorously-simulated coefficients of the far-field data, the red curves to the reconstructed far-field coefficients obtained with $\tilde{R}_{S}^{[1]}=\rho^{[1]}\phi^{-1},~\tilde{C}_{S}^{[1]}=c^{[1]}$ and the black curves to the reconstructed far-field coefficients obtained with $\tilde{C}_{D}^{[1]}$. The filling factor is $\phi=3/4$.  Case $h=0.00175~m$.}
  \label{honC1-10}
  \end{center}
\end{figure}

\begin{figure}[ptb]
\begin{center}
\includegraphics[width=12cm] {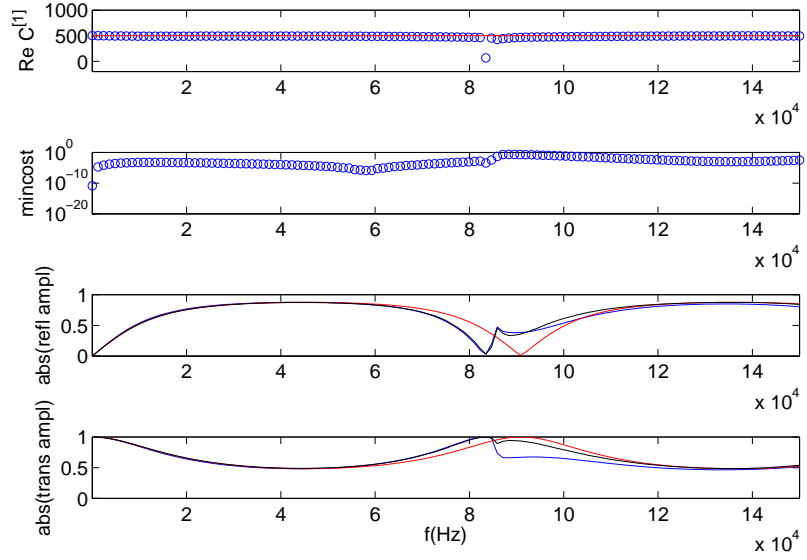}
 \caption{Same as fig. \ref{honC1-10} for case $h=0.00275~m$.}
  \label{honC1-20}
  \end{center}
\end{figure}

\begin{figure}[ptb]
\begin{center}
\includegraphics[width=12cm] {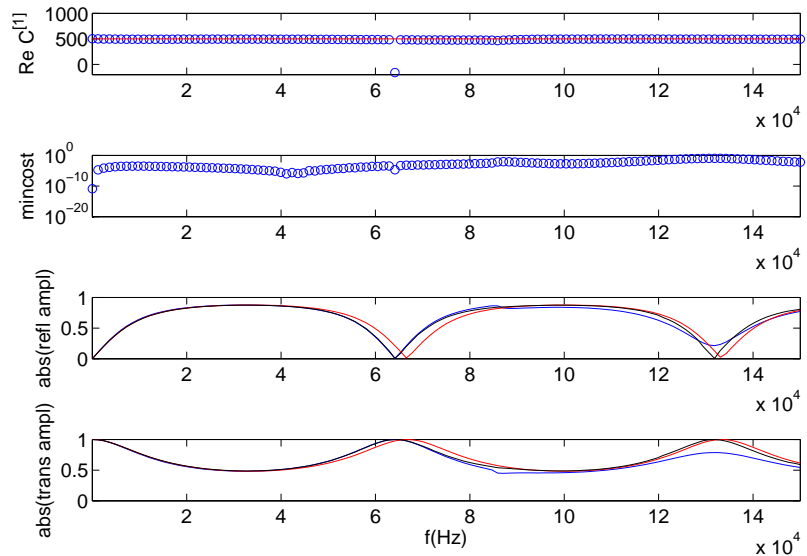}
 \caption{Same as fig. \ref{honC1-10} for case $h=0.00375~m$.}
  \label{honC1-30}
  \end{center}
\end{figure}

As written earlier, we would like the homogenized constitutive parameters to not depend on the given structure thickness=surrogate layer thickness=$h$. Figs.  \ref{honC1-10}-\ref{honC1-30} tell us how $\tilde{C}_{D}^{[1]}$ varies with $h$.
\newpage
Although the variation of $h$ produces variations in the far-field responses (see two lower panels in the figures) this has practically no effect on $\tilde{R}_{D}^{[1]}$ except in the close neighborhoods of the frequencies at which the far-field reflected response is nil, a situation that results in the inversion procedure giving rise to a slightly-resonant type of feature. The latter enables a rather good agreement between the far-field data and reconstructed far-field response which is better than that obtained via $\tilde{R}_{S}^{[1]}$. Nevertheless, the reconstructed surrogate layer far-field response obtained with $\tilde{R}_{S}^{[1]}$ is in good agreement with the simulated far-field data over a rather broad range of low frequencies.

Last  but not least we note that  $C_{D}^{[1]}=C_{S}^{[1]}=500~ms^{-1}$ (corresponding to $p_{C}=0$ over a broad range of frequencies.
\subsubsection{Retrieval of $\tilde{C}^{[1]}_{D}$  for  increasing $\theta^{i}$}
\begin{figure}[ht]
\begin{center}
\includegraphics[width=12cm] {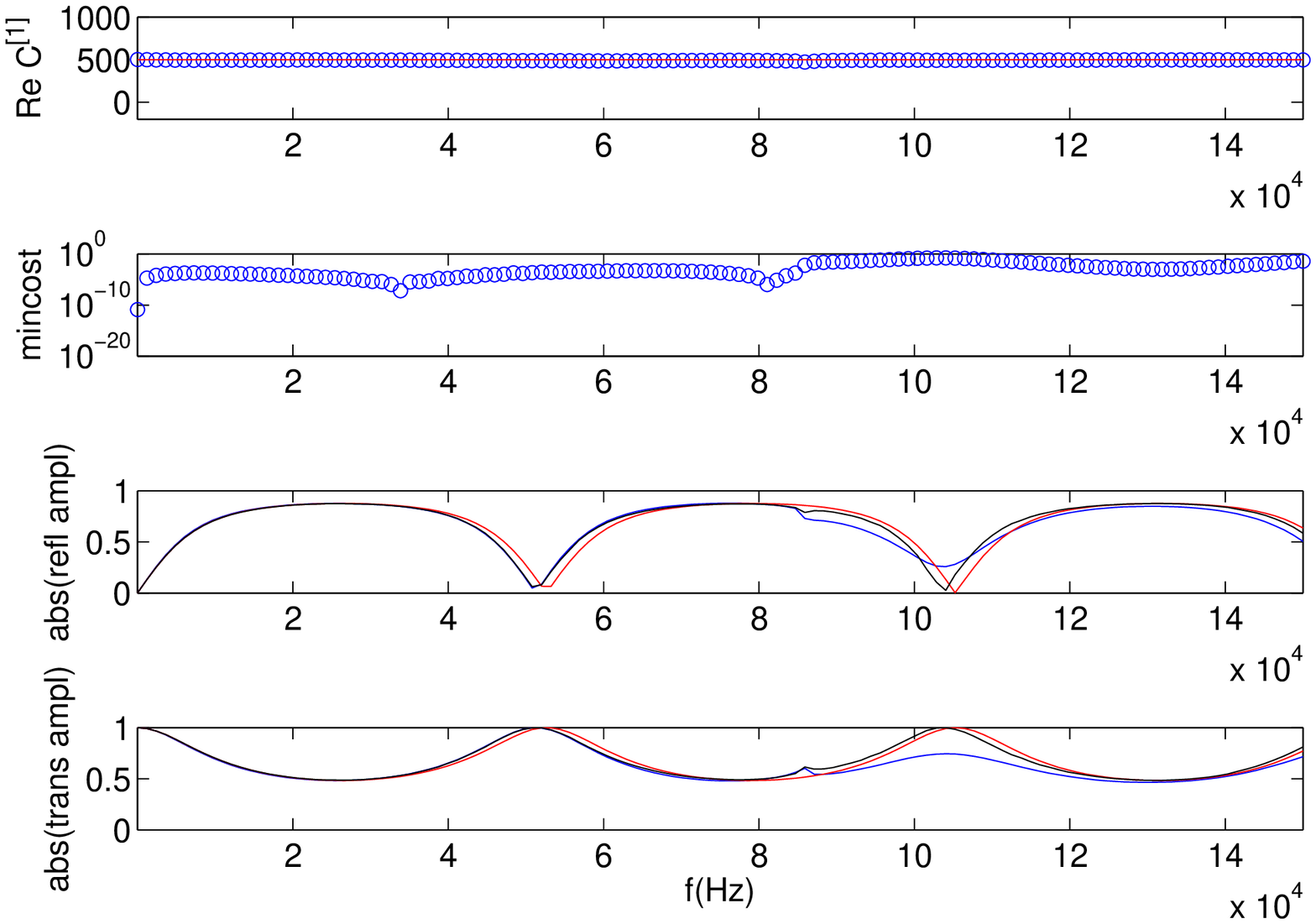}
 \caption{Retrieval of $\tilde{B}=\tilde{C}^{[1]}$. The red line corresponds to $\tilde{C}_{S}^{[1]}$ assuming $p_{C}=0$ and $p_{R}=-1$. The circles correspond to the retrieval of $\tilde{C}_{D}^{[1]}$ via cost functional $\kappa_{1}$. The blue curves in the third and fourth panels correspond to the rigorously-simulated coefficients of the far-field data, the red curves to the reconstructed far-field coefficients obtained with $\tilde{R}_{S}^{[1]}=\rho^{[1]}\phi^{-1},~\tilde{C}_{S}^{[1]}=c^{[1]}$ and the black curves to the reconstructed far-field coefficients obtained with $\tilde{C}_{D}^{[1]}$. The filling factor is $\phi=3/4$, $h=0.00475~m$.  Case $\theta^{i}=0^{\circ}$.}
  \label{thonC1-10}
  \end{center}
\end{figure}

\begin{figure}[ptb]
\begin{center}
\includegraphics[width=12cm] {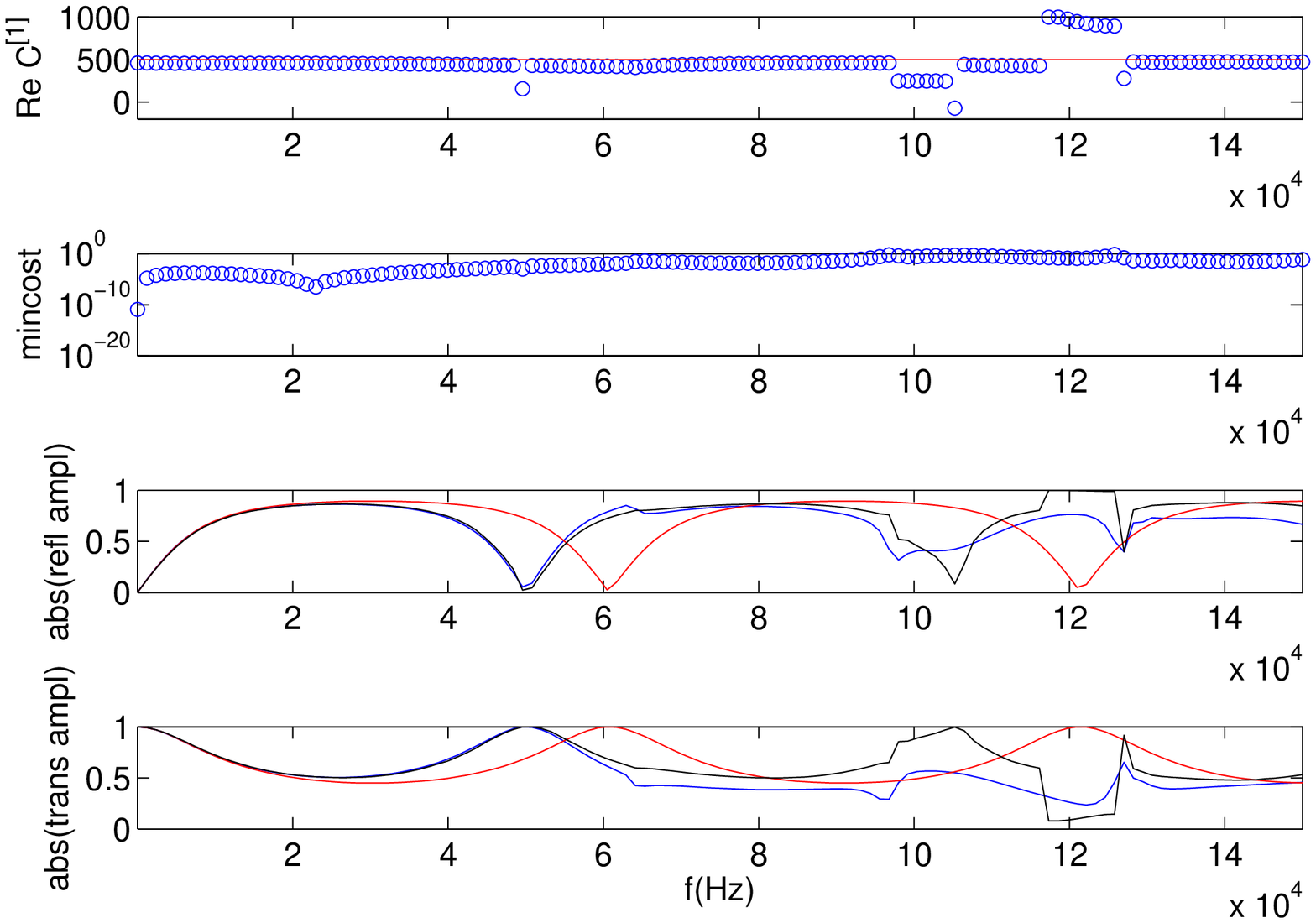}
 \caption{Same as fig. \ref{thonC1-10} for case $\theta^{i}=20^{\circ}$.}
  \label{thonC1-20}
  \end{center}
\end{figure}

\begin{figure}[ptb]
\begin{center}
\includegraphics[width=12cm] {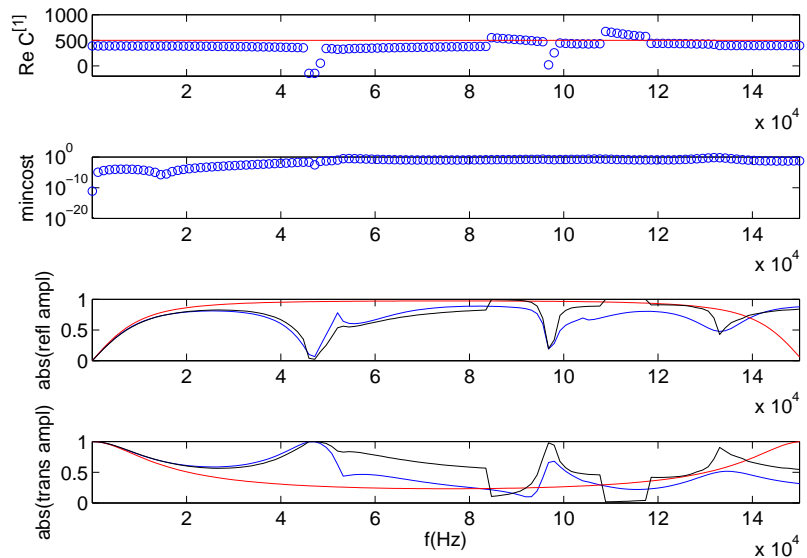}
 \caption{Same as fig. \ref{thonC1-10} for case $\theta^{i}=40^{\circ}$.}
  \label{thonC1-30}
  \end{center}
\end{figure}

As indicated earlier, we would like the homogenized constitutive parameters to not depend on the incident angle $\theta^{i}$.  Figs.  \ref{thonC1-10}-\ref{thonC1-30} tell us how $\tilde{C}_{D}^{[1]}$ varies with $\theta^{i}$.
\newpage
Although the increase of $\theta^{i}$ produces  increasing discrepancies between the far-field data and the reconstructed  far-field responses (see two lower panels in the figures) this is not very noticeable in the variations of $C_{D}^{[1]}$ with this external parameter except for what appears as a small decrease of the retrieved average velocity and the anomalous features that become more pronounced with increasing $\theta^{i}$. A remarkable feature of these figures is that they show how $C_{D}^{[1]}$ results in much better agreement than does $C_{S}^{[1]}$ between the far-field data and the reconstructed far-field response, which fact again points to the "curative virtue" of dispersion introduced by the retrieval method. However, the curative power of this  'induced dispersion' \cite{wi16a,wi16b,wi16c} has its limits as seen in fig.  \ref{thonC1-30}.

Thus, it is not possible to conclude that our homogenized layer responds independently of $\theta^{i}$, this being due to the fact that our dynamic homogenization recipe does not take into account the external parameter $\theta^{i}$.
\subsubsection{Retrieval of $\tilde{C}^{[1]}_{D}$  for constant $d$ and  increasing $w$}
\begin{figure}[ht]
\begin{center}
\includegraphics[width=12cm] {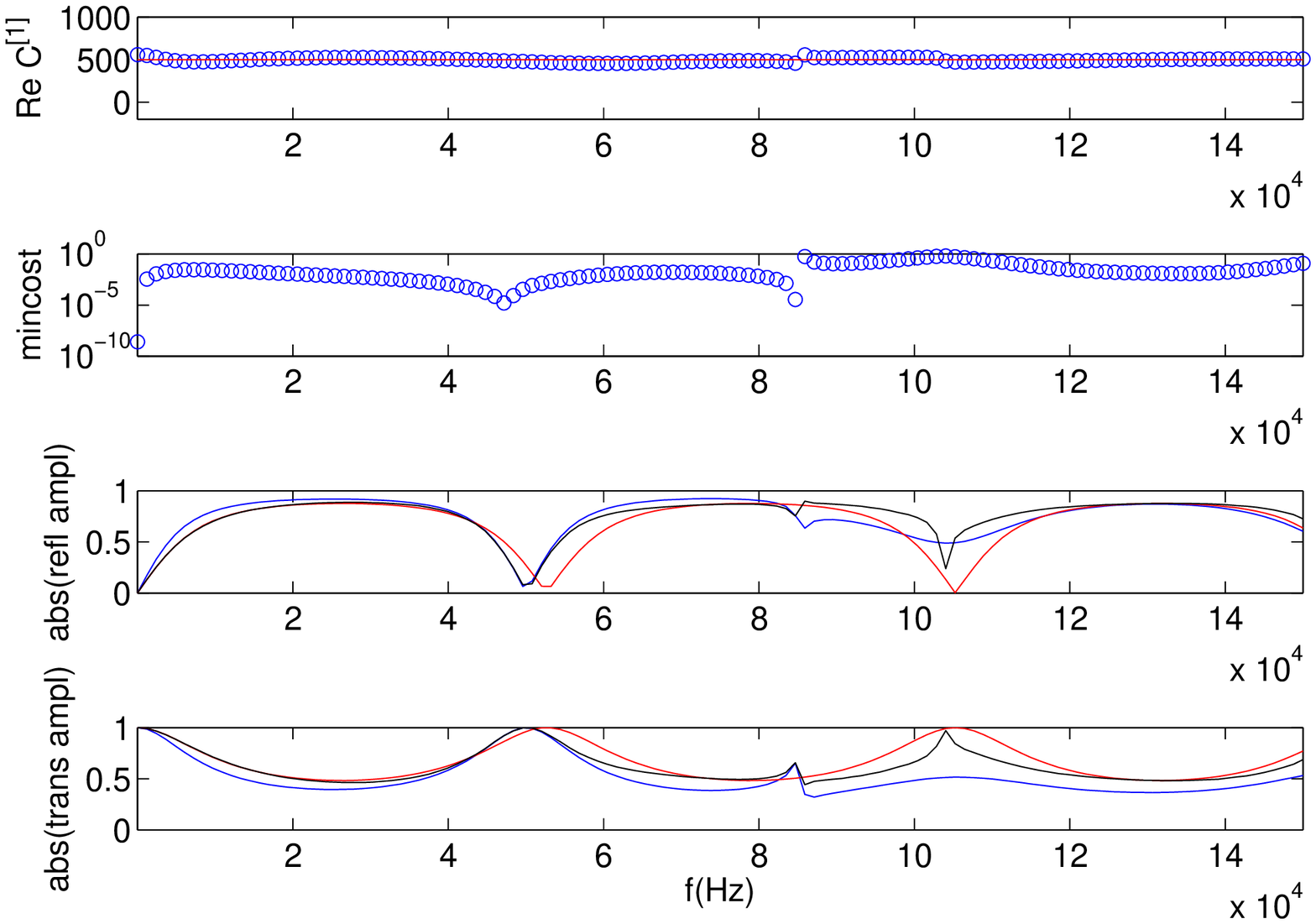}
 \caption{Retrieval of $\tilde{B}=\tilde{C}^{[1]}$. The red line corresponds to $\tilde{C}_{S}^{[1]}$ assuming $p_{C}=0$ and $p_{R}=-1$. The circles correspond to the retrieval of $\tilde{C}_{D}^{[1]}$ via cost functional $\kappa_{1}$. The blue curves in the third and fourth panels correspond to the rigorously-simulated coefficients of the far-field data, the red curves to the reconstructed far-field coefficients obtained with $\tilde{R}_{S}^{[1]}=\rho^{[1]}\phi^{-1},~\tilde{C}_{S}^{[1]}=c^{[1]}$  and the black curves to the reconstructed far-field coefficients obtained with $\tilde{C}_{D}^{[1]}$. $h=0.00475~m$. $\theta^{i}=0^{\circ}$.  Case $w=0.0024~m$ ($\phi=0.6$).}
  \label{wonC1-10}
  \end{center}
\end{figure}

\begin{figure}[ptb]
\begin{center}
\includegraphics[width=12cm] {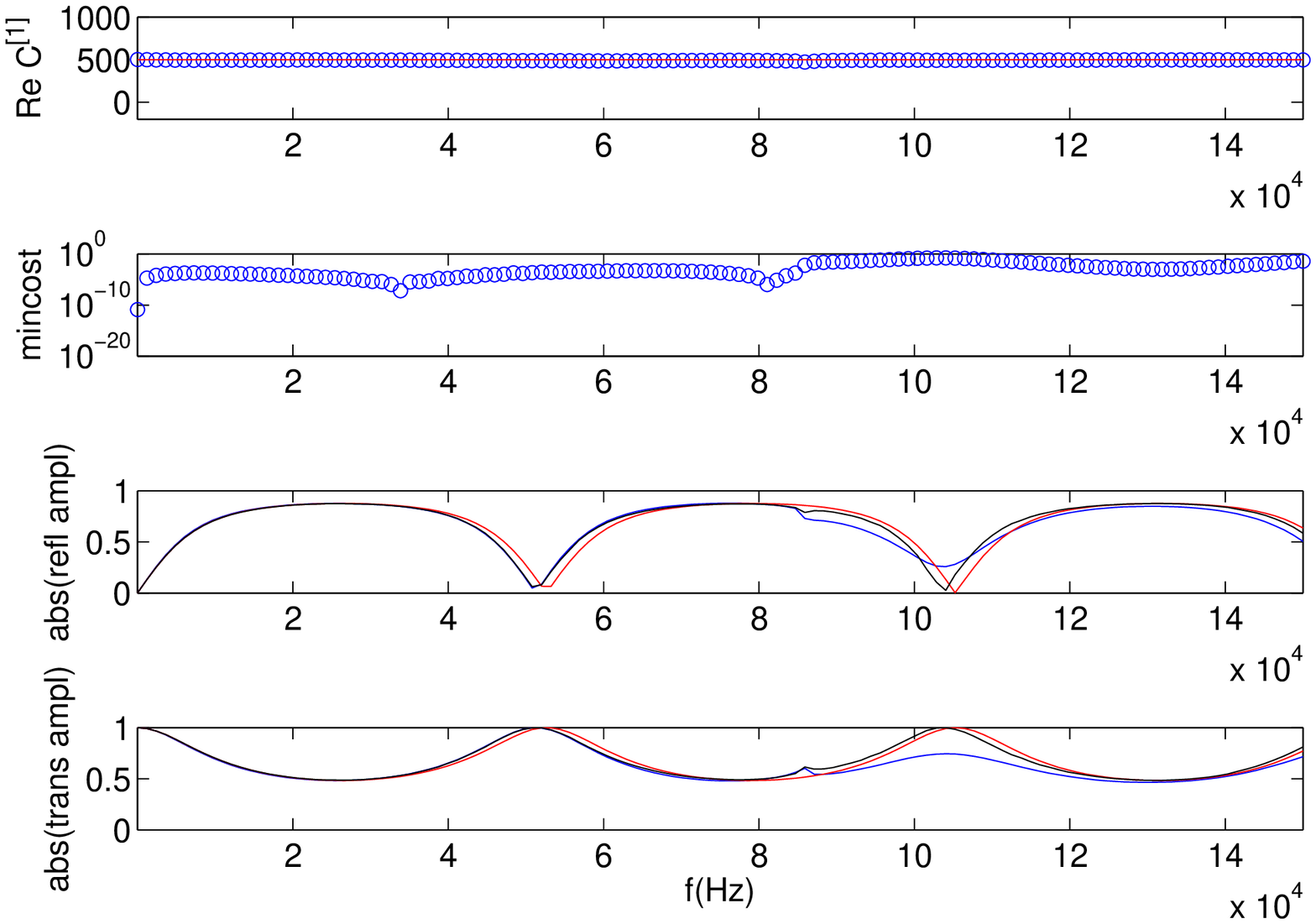}
 \caption{Same as fig. \ref{wonC1-10} for case $w=0.0030~m$ ($\phi=0.75$).}
  \label{wonC1-20}
  \end{center}
\end{figure}

\begin{figure}[ptb]
\begin{center}
\includegraphics[width=12cm] {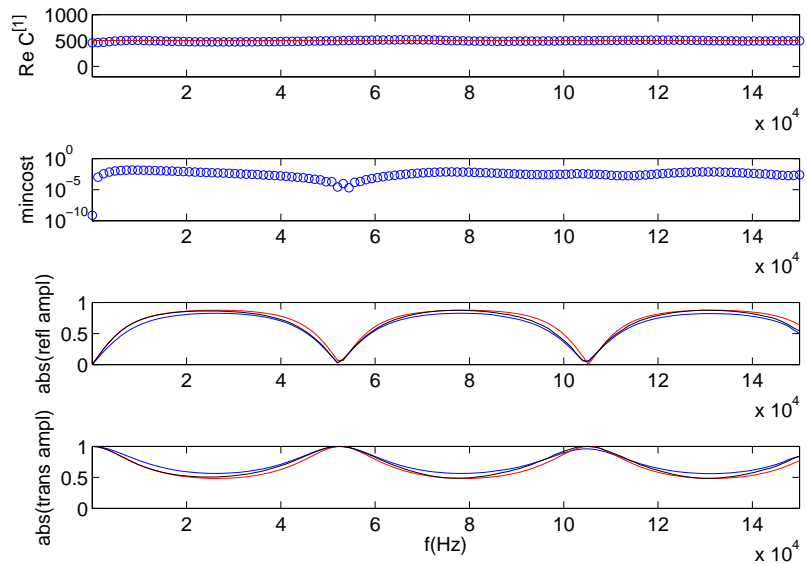}
 \caption{Same as fig. \ref{wonC1-10} for case $w=0.0036~m$ ($\phi=0.90$).}
  \label{wonC1-30}
  \end{center}
\end{figure}

There exist good reasons to believe \cite{wi18} that our homogenization scheme works less well for smaller filling factors $\phi$. To address this issue, we kept $d$ constant ($=0.004~m$ as usual) and varied $w$, which gave rise to the results in figs.  \ref{wonC1-10}-\ref{wonC1-30}.
\newpage
This is substantiated by the results in these figures wherein we note once again the curative effect of taking into account the dispersion of the retrieved density for the reconstructed far-field response, especially at low $\phi$.
\subsection{Total transmission}
There have appeared many examples of total transmission at certain frequencies in the previous figures, but  no particular attention was given to this phenomenon because it seemed to be natural for transmission gratings having rather {\it wide} spaces (i.e., $\phi=w/d \ge 3/5$) between successive rigid blocks.

In \cite{ll07} the same sort of phenomenon was predicted numerically (by finite elements) and substantiated experimentally, for  an acoustic wave striking a grating similar to ours having {\it narrow} spaces (i.e., $\phi=w/d=1/9$) between its successive rigid blocks, and qualified as being 'extraordinary' (acoustic transmission, EAT for short) because it would not be expected that such an almost-opaque structure let so much sound penetrate it.

\begin{figure}[ht]
\begin{center}
\includegraphics[width=12cm] {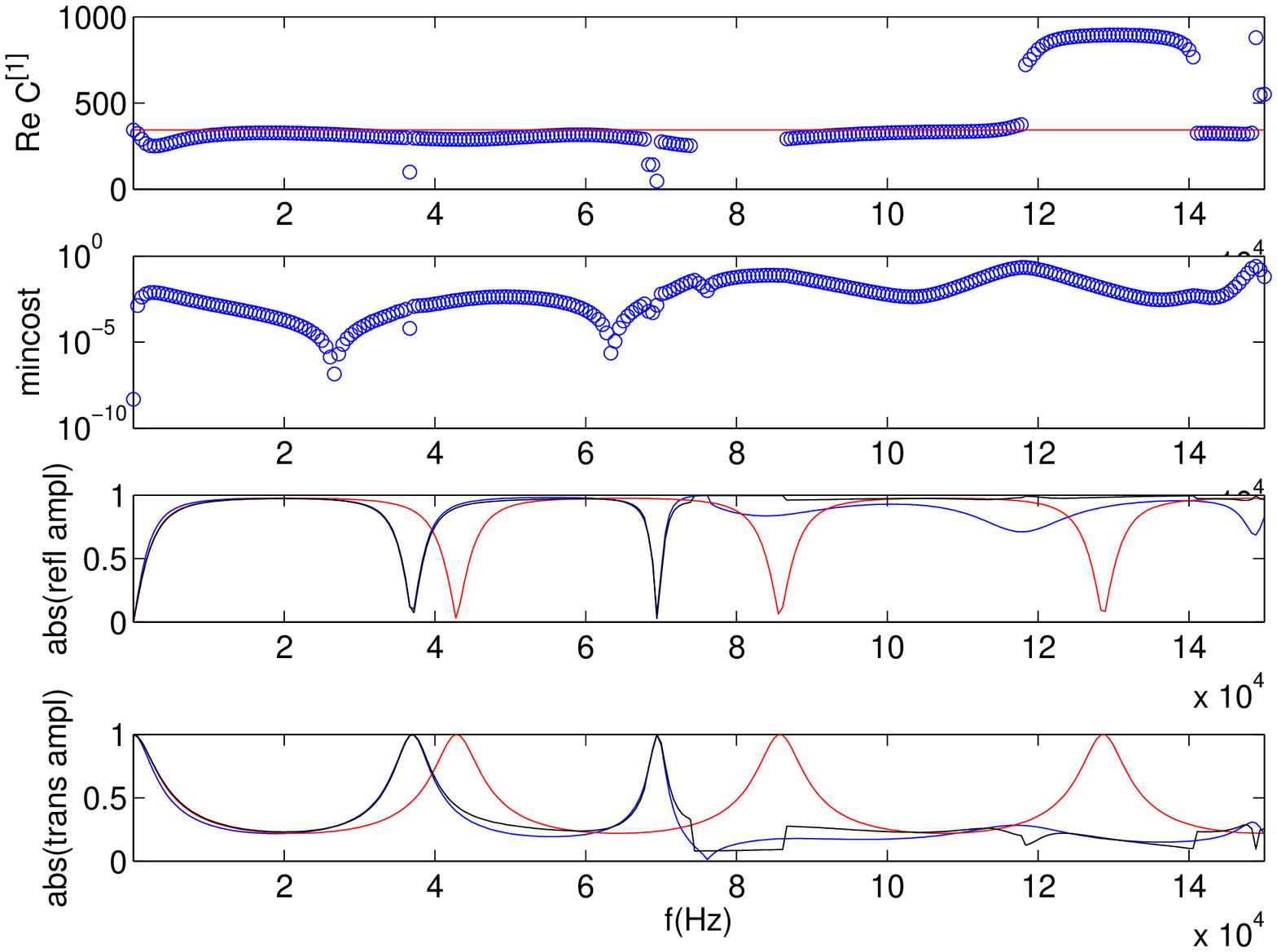}
 \caption{Retrieval of $\tilde{B}=\tilde{C}^{[1]}$. The red line corresponds to $\tilde{C}_{S}^{[1]}$. The circles correspond to the retrieval of $\tilde{C}_{D}^{[1]}$ via cost functional $\kappa_{1}$. The blue curves in the third and fourth panels correspond to the rigorously-simulated coefficients of the far-field data, the red curves to the reconstructed far-field coefficients  obtained with $\tilde{R}_{S}^{[1]},~\tilde{C}_{S}^{[1]}$  and the black curves to the reconstructed far-field coefficients obtained with $\tilde{C}_{D}^{[1]}$. $d=0.0045~m$, $w=0.0005~m$ (i.e., $\phi=1/9$), $h=0.004~m$. $\theta^{i}=0^{\circ}$.  $p_{C}=0$, (i.e., $\tilde{C}_{S}^{[1]}=c^{[1]}=343~ms^{-1}$), $p_{R}=-1$ (i.e., $\tilde{R}_{S}^{[1]}=\rho^{[1]}\phi^{-1}=10.8~Kgm^{-1}$). $R^{[1]}=\tilde{R}_{S}^{[1]}$ was assumed for the retrieval of $\tilde{C}_{D}^{[1]}$}.
  \label{tt-10}
  \end{center}
\end{figure}

\begin{figure}[ht]
\begin{center}
\includegraphics[width=12cm] {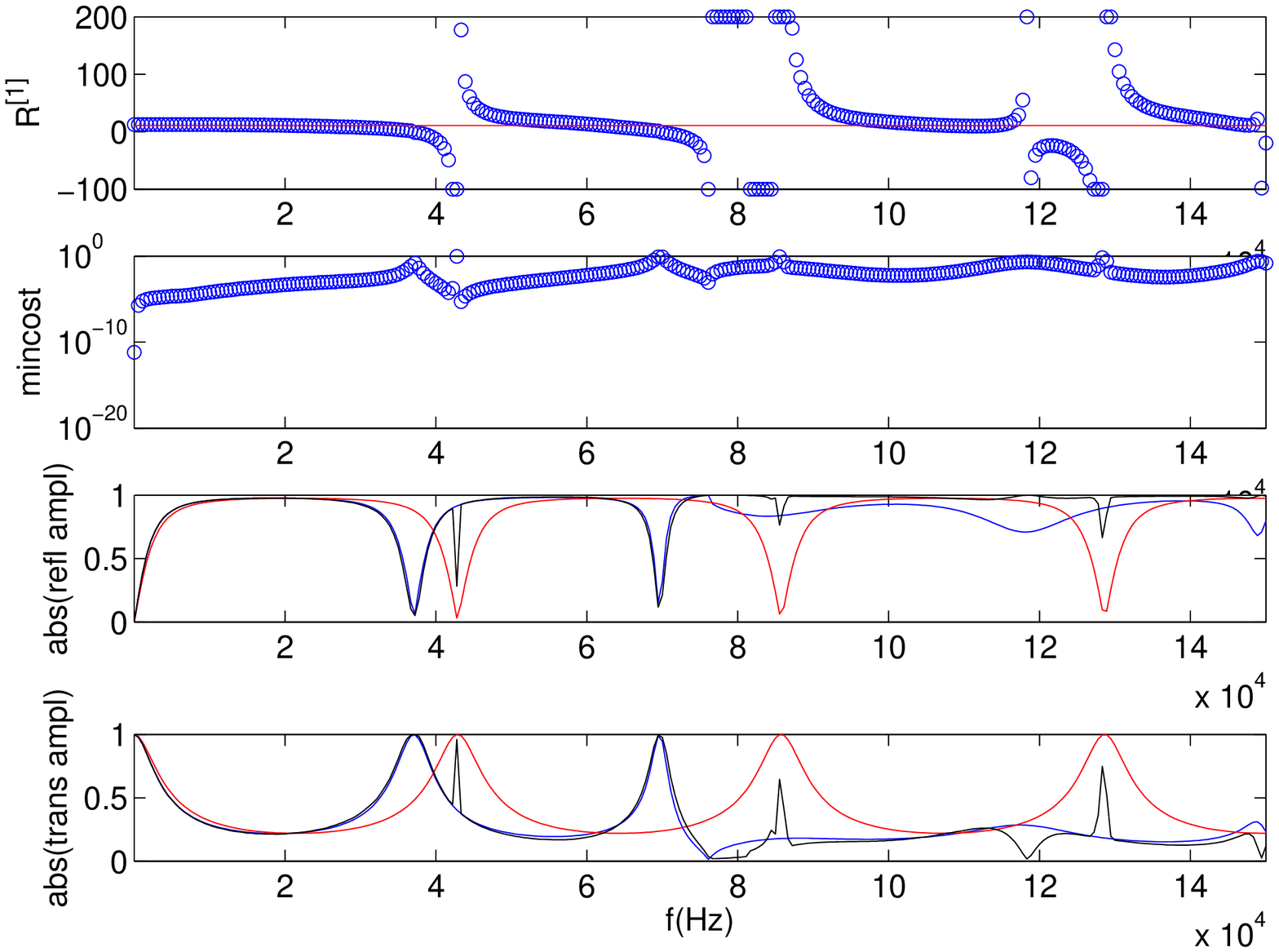}
 \caption{Retrieval of $\tilde{B}=\tilde{R}^{[1]}$. The red line corresponds to $\tilde{C}_{S}^{[1]}$. The circles correspond to the retrieval of $\tilde{C}_{D}^{[1]}$ via cost functional $\kappa_{1}$. The blue curves in the third and fourth panels correspond to the rigorously-simulated coefficients of the far-field data, the red curves to the reconstructed far-field coefficients obtained with $\tilde{R}_{S}^{[1]}$ and the black curves to the reconstructed far-field coefficients obtained with $\tilde{R}_{D}^{[1]}$. $d=0.0045~m$, $w=0.0005~m$ (i.e., $\phi=1/9$), $h=0.004~m$. $\theta^{i}=0^{\circ}$.  $p_{C}=0$, (i.e., $\tilde{C}_{S}^{[1]}=c^{[1]}=343~ms^{-1}$), $p_{R}=-1$ (i.e., $\tilde{R}_{S}^{[1]}=\rho^{[1]}\phi^{-1}=10.8~Kgm^{-1}$). $C^{[1]}=\tilde{C}_{S}^{[1]}$ was assumed for the retrieval of $\tilde{R}_{D}^{[1]}$}
  \label{tt-20}
  \end{center}
\end{figure}

\begin{figure}[ht]
\begin{center}
\includegraphics[width=12cm] {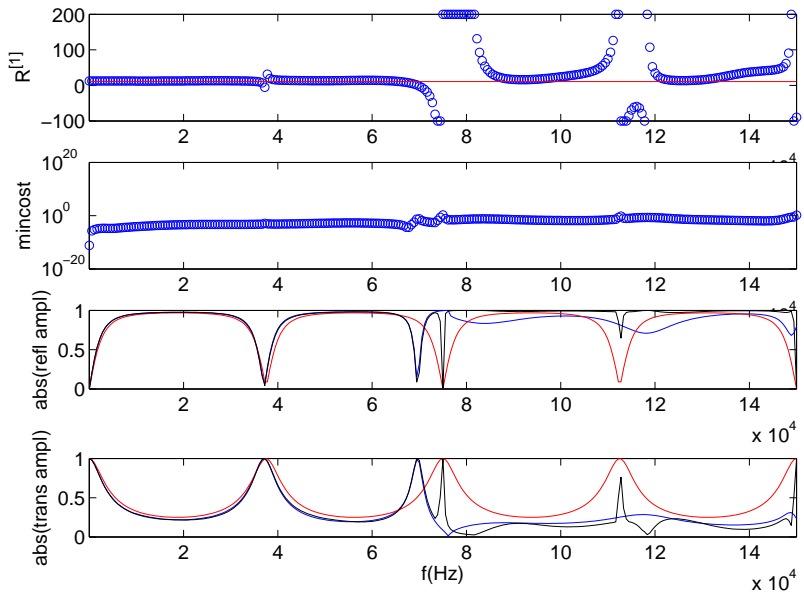}
 \caption{Everything the same as in fig. \ref{tt-20} except that $C^{[1]}=\tilde{C}_{S}^{[1]}$ is taken to be $300~ms^{-1}$, i.e., lower than in the previous two figures.}
  \label{tt-30}
  \end{center}
\end{figure}

We wondered whether our DD-SOV method could  produce the same sort of prediction, and if so, what the parameters of the layer surrogate would result from the retrieval scheme. Unfortunately, the authors of \cite{ll07} forgot to mention what the acoustic parameters of their site were, but we guess that it was air at $20^{\circ}C$ for which the constitutive parameters are $c^{[j]}=343~ms^{-1}~;~j=0,1,2$ and $\rho^{[j]}=1.2~Kgm^{-3}~;~j=0,1,2$. The other parameters were taken from \cite{ll07} and our computations thus led to the results exhibited in figs. \ref{tt-10}-\ref{tt-30}.
\newpage
In the lowermost panel of fig. \ref{tt-10} (relative to the retrieval of the effective velocity) we observe the same two total transmission peaks as those shown in \cite{ll07}. In the two lowermost panels we find good agreement between the simulated data  and the reconstructed far-field responses from the retrieved dynamic parameters but poor agreement between the simulated data and the reconstructed far-field responses using the static parameters. On the basis of what we observe in the uppermost panel we guess that this discrepancy could be due, at least at low frequencies, to an overestimation of $\tilde{C}_{S}^{[1]}$. We shall return to this issue in the discussion of the next two figures.
\newpage
In the lowermost panel of fig. \ref{tt-20} (relative to the retrieval of the effective mass density) we again observe the same two total transmission peaks as those shown in \cite{ll07}. In the two lowermost panels we again find good agreement between the simulated data  and the reconstructed far-field responses from the retrieved dynamic parameter (except for the existence of some small peaks probably due to the retrieval scheme overdoing its curative role) but poor agreement between the simulated data and the reconstructed far-field responses from the static parameters. On the basis of what we observe in the uppermost panel of the previous figure we guess that this discrepancy could be due, at least at low frequencies, to an overestimation of $C^{[1]}=\tilde{C}_{S}^{[1]}$. We shall test this hypothesis in the next figure.
\newpage
In the lowermost panel of fig. \ref{tt-30} (relative, once more, to the retrieval of the effective velocity) we again observe the same two total transmission peaks as those shown in \cite{ll07}. In the two lowermost panels we again find good agreement between the simulated data  and the reconstructed far-field responses from the retrieved dynamic parameter (except for the existence of extraneous  peaks which occur at higher frequency than previously and are probably due to the retrieval scheme overdoing its curative role) and also good agreement between the simulated data and the reconstructed far-field responses from the static parameters in the low frequency (up till $\sim 65~KHz$) region.

This shows that the static homogenized layer model is not valid as such for small $\phi$, but can be corrected (up till $\sim 65~KHz$) quite simply by a change of $p_{C}$ (here amounting to a reduction of $\tilde{C}_{S}^{[1]}$).

We have thus substantiated the finding in \cite{ll07} of total transmission at two distinct low frequencies and found, in addition, that the grating structure with very narrow spaces between successive blocks responds to a normally-incident plane acoustic wave as does a homogenous layer having constitutive parameters $\tilde{C}_{S}=300~ms^{-1}$ and $\tilde{R}_{S}=10.8~Kgm^{-3}$. This combination of acoustic constitutive parameters does not, to our knowledge, correspond to any existing macroscopically-homogeneous medium.
\clearpage
\newpage
\section{Conclusions}
By inversion  of its far-field response to acoustic body wave solicitations we have found that the periodic array  of rigid rectangular block structure  behaves much like a homogeneous  layer (surrogate):\\
 (i) whose dynamic (and static) effective mass density is equal to the product of the mass density (assumed to be non dispersive) of the generic block of the periodic structure with the inverse of the filling factor (proportion of the width of  the space between successive blocks to the period of the structure),\\
 (ii) whose dynamic  (and static)  velocity (real) is equal to the velocity (assumed to be real and non-dispersive) of the medium located  between successive blocks,\\
 (iii) all of whose constitutive properties do not depend on frequency (i.e., the surrogate layer is temporally non-dispersive).\\\\
  This conclusion holds:\\
   (a) for low frequencies (i.e., less than $\sim 50~KHz$ for a structure whose period is $0.004~m$) other than those at which a Wood anomaly can  be excited in the given structure,\\
   (b) for filling factors that are larger than approximately 0.5,\\
   (c) whatever be the  incident angles of the solicitation (although the larger the incident angle, the smaller the low-frequency bandwidth of equivalence) other than those at which a Wood anomaly is excited,\\
   (d) whatever be the block thickness (equal to the  surrogate layer thickness) provided it is not too small, and \\
   (e) irrespective of the site,\\
   (f) even for the phenomenon of total transmission \cite{ll07} when the spaces between blocks are very narrow).\\\\
   This means that the surrogate layer model with properties (i)-(iii) (which we call SLM) is a true effective medium under conditions (a)-(f), with mathematically-explicit and simple, constitutive properties.

When any of the conditions (a)-(c) are not satisfied, it was found that the retrieved constitutive parameters depart from those of the SLM all the more so the smaller is the filling factor and the closer is the frequency and incident angle to those at which a Wood anomaly is excited. This raises the question of the meaning of what, notably the constitutive parameters being a function of frequency, is predicted by the response inversion procedure model when any of conditions (a)-(c) are not satisfied. The answer is what may be termed 'apparent dispersion' \cite{mo10,wi16c} or 'induced dispersion', this meaning that the dispersion, which was assumed to be absent in each component of the original configuration, is a result of the surrogate (layer) model  telling a story that is different from the one told by the original (period structure) model concerning its response to the solicitation (the same is true in the first attempts at dynamic homogenization \cite{ua49,wt61}),  this discordance being particularly-noticeable when any of conditions (a)-(c) are violated.  This does not mean that the  retrieved dispersive constitutive parameters (we term them as being obtained by the DSLM) are erroneous since (in the sense of inversion) they correspond to a minimum of the discrepancy functional at these frequencies, incident angles and for these filling fractions, but the value of the cost at these minima is usually larger than when conditions (a)-(c) are satisfied. Thus, our (DSLM) dynamic homogenization scheme seems to be doing the right thing, i.e., compensating, by inducing dispersion, for the aforementioned discordance \cite{wi16b} in order to match, as closely as possible, the (far, but probably also near) fields of the surrogate layer to those of the periodic structure. This apparently-advantageous aspect  is offset by the fact that it becomes difficult to extract simple relations for the DSLM constitutive properties at frequencies at which anomalous response is observed or, more, generally, at higher frequencies.

In this respect, it is instructive to compare the results of our inversion procedure  to those of a multiple-scales (theoretical) construction of an effective medium. For this, we refer to the publication \cite{fb05} of Felbacq and Bouchitt\'e  who obtain the effective medium corresponding to the original medium composed of three superposed gratings of square dielectric cylinders solicited by an electromagnetic plane body wave. The authors of this publication find a simple expression for the effective permeability (their eq. (31)) which is clearly-dispersive whereas the effective permittivity takes on a constant value for all frequencies. Moreover, in their fig. 3, they compare the transmission  of the original grating structure to that of the homogenized layer and find that the two compare very well except in the neighborhood of  a frequency at which the grating structure gives rise to an anomalous feature whereas this feature is absent in the transmission associated with the homogenized layer. We carried out the same sort of comparison, but now concerning the far-reflected and transmitted fields  of the periodic structure and its layer surrogate.
\begin{figure}[ht]
\begin{center}
\includegraphics[width=12cm] {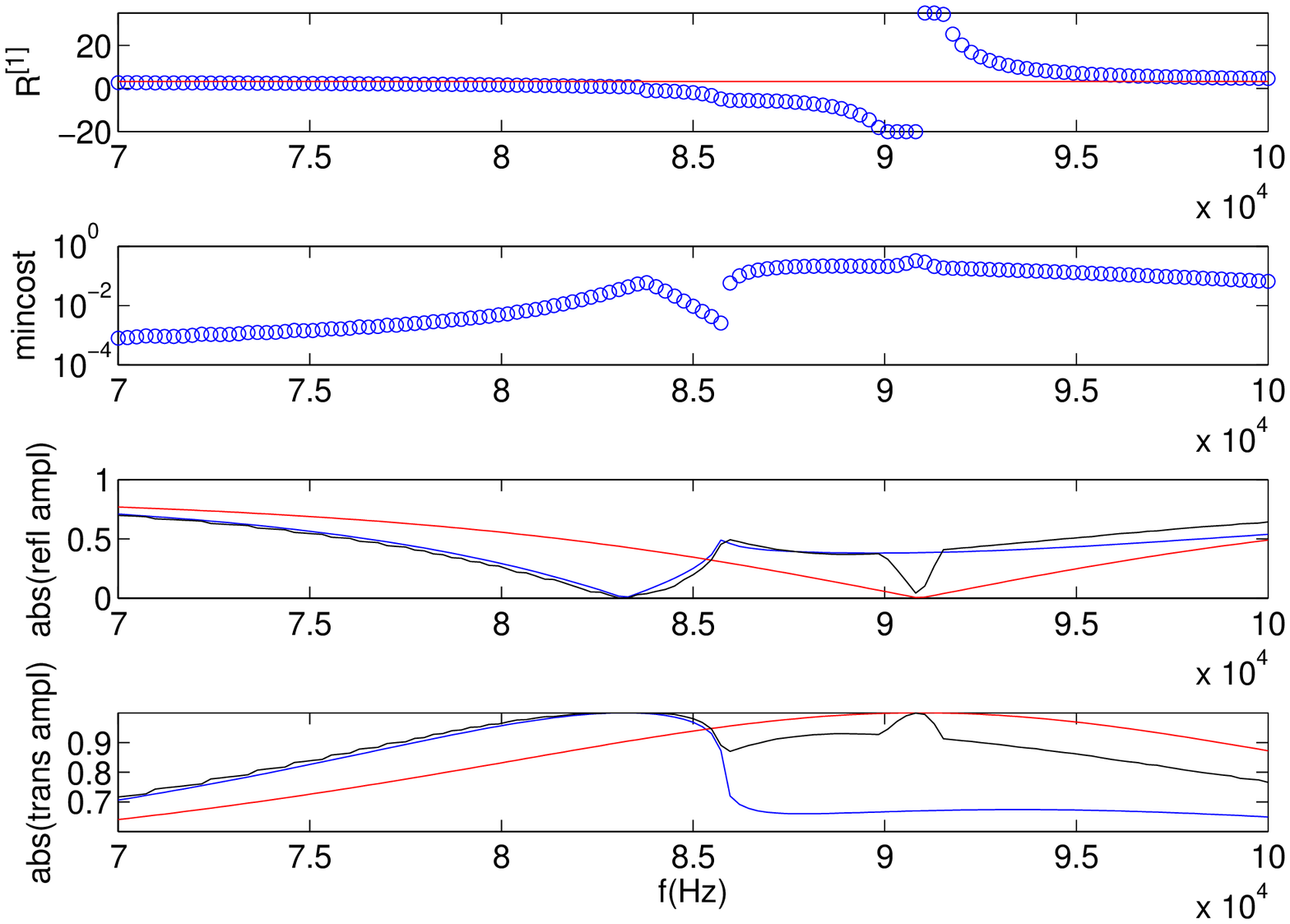}
 \caption{Zoom of fig. \ref{honR1-10}. Retrieval of $\tilde{B}=\tilde{R}^{[1]}$. The red line  corresponds to $\tilde{R}_{S}^{[1]}$ assuming $p_{C}=0$ and $p_{R}=-1$. The circles correspond to the retrieval of $\tilde{R}_{D}^{[1]}$ via cost functional $\kappa_{1}$. The blue curves in the third and fourth panels correspond to the rigorously-simulated coefficients of the far-field data, the red curves to the reconstructed far-field coefficients obtained with $\tilde{R}_{S}^{[1]}$ and the black curves to the reconstructed far-field coefficients obtained with $\tilde{R}_{D}^{[1]}$. The filling factor is $\phi=3/4$. $C^{[1]}=c^{[1]}$. Case $h=0.00275~m$.}
  \label{nf-10}
  \end{center}
\end{figure}
 We note in  fig. \ref{nf-10} that: (1) the SLM  prediction (red curves in the two lowest panels) compares fairly-well, as expected, with the given structure response (blue curves in the two lowest panels)  at low frequencies, but {\em does not account} for the Wood anomalous feature  near $86~KHz$, this being similar to what was found in fig. 3 of \cite{fb05}, (2) the DSLM prediction (black curves in the two lowest panels) compares at least as well as the SLM at low frequencies, but {\em does account}  for the anomalous feature near $86~KHz$.  The missing element in the SLM is then clearly the relatively-high frequency dispersive features  that appear in the DSLM, which we have not  described in explicit, mathematical terms. Such an element must also be absent in the multiscale construction of permittivity and permeability of \cite{fb05} for the same reason.

Despite this shortcoming, the lower two panels of figs. \ref{honR1-10} and \ref{nf-10} indicate (at least for this particular  solicitation and periodic structure for which the Wood anomaly contribution to the response spectrum is rather weak) that the far-field transfer functions  are rather well-predicted by both the SLM and DSLM.

The 'reality' of a 2D periodic structure and plane wave solicitation due to very distant sources is, of course, doubtful, which is why the method employed herein must be extended to non-periodic structures with some disorder in its various geometric and compositional parameters, as well as to solicitations  due to  sources that are nearer the scattering structure. It can be anticipated that something similar to our SLM will be able to account for some of the salient features of the  response of such more realistic scattering structures and solicitations.
%

\end{document}